\documentclass{egpubl}
\usepackage{eurovis2024}

\SpecialIssuePaper         

 \STAREurovis   

\CGFccby

\usepackage[T1]{fontenc}
\usepackage{dfadobe}  

\usepackage{cite}  
\BibtexOrBiblatex
\electronicVersion
\PrintedOrElectronic
\ifpdf \usepackage[pdftex]{graphicx} \pdfcompresslevel=9
\else \usepackage[dvips]{graphicx} \fi
\usepackage{multirow}

\usepackage{egweblnk}

\usepackage{booktabs}
\usepackage{multirow}
\usepackage{tabularx}
\usepackage[normalem]{ulem}
\useunder{\uline}{\ul}{}
\usepackage[nameinlink]{cleveref}
\usepackage{microtype}
\usepackage[backgroundcolor=yellow!40]{todonotes} 
\usepackage{xcolor}

\newif\ifShowRevisions
\ShowRevisionsfalse

\newif\ifShowDeletions
\ShowDeletionsfalse

\ifShowRevisions
\newcommand{\new}[1]{\textcolor{teal}{#1}}
\newcommand{\revized}[1]{\textcolor{purple}{#1}}
\else
\newcommand{\new}[1]{#1}
\newcommand{\revized}[1]{#1}
\fi

\usepackage[normalem]{ulem}
\ifShowDeletions
\newcommand{\deleted}[1]{\textcolor{red}{\sout{#1}}}
\else
\newcommand{\deleted}[1]{}
\fi

\title[A State-of-the-Art Report on the Integration of Sonification and Visualization]
      {Open Your Ears \revized{and} Take a Look: \\ A State-of-the-Art Report on the Integration of Sonification and Visualization}

\author[K. Enge et al.]
{\parbox{\textwidth}{\centering 
    K.\,Enge$^{1,2}$\orcid{0000-0002-5456-1140},
    E.\,Elmquist$^{3}$\orcid{0000-0001-5874-6356},
    V.\,Caiola$^{4}$\orcid{0000-0002-4155-1234},
    N.\,Rönnberg$^{3}$\orcid{0000-0002-1334-0624},
    A.\,Rind$^{1}$\orcid{0000-0001-8788-4600},
    M.\,Iber$^{1}$\orcid{0000-0002-5929-8716}, \\
    S.\,Lenzi$^{5,6}$\orcid{0000-0001-5534-0596},
    F.\,Lan$^{7}$\orcid{0000-0002-8237-5919},
    R.\,Höldrich$^{2}$\orcid{0000-0002-6887-6637},
    and
    W.\,Aigner$^{1}$\orcid{0000-0001-5762-1869}
}
        \\
{\parbox{\textwidth}{\centering 
    $^1$Institute of Creative\textbackslash Media/Technologies (IC\textbackslash M/T), St. P{\"o}lten University of Applied Sciences, St. P{\"o}lten, Austria \\
    $^2$Institute of Electronic Music and Acoustics (IEM), University of Music and Performing Arts Graz, Graz, Austria\\
    $^3$Division for Media and Information Technology (MIT), Linköping University, Linköping, Sweden \\
    $^4$School of Creative Media, City University of Hong Kong, Kowloon, Hong Kong\\
    $^5$Ikerbasque Basque Foundation for Science, Bilbao, Spain\\  
    $^6$Faculty of Engineering, Universidad de Deusto, Bilbao, Spain\\
    $^7$Scientific Computing and Imaging Institute (SCI), University of Utah, Salt Lake City, USA
}}
}

\begin{document}


\maketitle

\begin{abstract}

The research communities studying visualization and sonification for data display and analysis share exceptionally similar goals, essentially making data of any kind interpretable to humans. One community does so by using visual representations of data, and the other community employs auditory (non-speech) representations of data. While the two communities have a lot in common, they developed mostly in parallel over the course of the last few decades. With this STAR, we discuss a collection of work that bridges the borders of the two communities, hence a collection of work that aims to integrate the two techniques into one form of audiovisual display, which we argue to be ``more than the sum of the two.''
We introduce and motivate a classification system applicable to such audiovisual displays and categorize a corpus of 57 academic publications that appeared between 2011 and 2023 in categories such as reading level, dataset type, or evaluation system, to mention a few. The corpus also enables a meta-analysis of the field, including regularly occurring design patterns such as type of visualization and sonification techniques, or the use of visual and auditory channels, \new{showing an overall diverse field with different designs}. \revized{An analysis of a co-author network of the field shows individual teams without many interconnections}. The body of work covered in this STAR also relates to three adjacent topics: audiovisual monitoring, accessibility, and audiovisual data art. These three topics are discussed individually in addition to the systematically conducted part of this research. The findings of this report may be used by researchers from both fields to understand the potentials and challenges of such integrated designs while hopefully inspiring them to collaborate with experts from the respective other field.


\end{abstract}


\section{Introduction} 

Over the course of the last few decades, two research communities have developed largely in parallel: one studying data visualization and one studying sonification. While the visualization community is primarily interested in ``the use of computer-supported, interactive, visual representations of abstract data to amplify cognition''~\cite{card_1999_readings}, the sonification community studies ``the use of non-speech audio to convey information''~\cite{kramer_1999_report}. To this day, the communities seem to be largely \revized{disjoint}, despite their shared goals promising fruitful collaboration. Both the theoretical cross-pollination~\cite{enge_2023_unified,caiola_2022_AudiovisualSonificationsDesign} and the practical integration and combination of sonification and visualization offer potential for interesting research outcomes. Therefore, with this state-of-the-art report (STAR), we want to shed light on audiovisual display idioms that systematically integrate data visualization and sonification. Informed by the definition of the visual idiom by Munzner~\cite{munzner_visualization_2015}, we think of an \emph{audiovisual display idiom} as ``a distinct approach to creating and manipulating audiovisual representations of data.''

In our daily lives, \revized{we perceive our surroundings in inherently multimodal and transmodal ways}. We see, and we hear, we read books, and we listen to music. We use our senses to understand and explore the world around us. Although we are multisensorial beings, the predominant data analysis idioms are unimodal, often using visualization only. Sight and hearing are inherently different, with different strengths and challenges, and are most probably suitable for different approaches in data representation. \revized{Human} visual perception can look upon a visual representation in a non-linear fashion, and the use of different types of graphs, charts, and other visual formats can reveal patterns, correlations, and trends in data that are often not as noticeable in numerical form.
Visual representations of data can also be experienced as more engaging and memorable to a user compared to tables with numbers~\cite{franconeri2021science}.
Sonification, on the other hand, exploits the excellent ability of the human auditory system to recognize temporal changes and patterns. It is, therefore, useful when displaying complex patterns, such as changes in time and warnings for immediate action. In real-time environments, sonification allows a controller to perceive information without the constant monitoring of visual displays. Sonification also enables the communication of data and information for visually impaired individuals~\cite{walker2010universal}. 

The \revized{human} visual perception has some challenges that can be supported by sonification, and similarly, the auditory system has other challenges that, in turn, can be supported by visualization. Therefore, we believe that a well-designed audiovisual representation can be more than the mere sum of a visual and an auditory representation. To put this into practice, we refer to a typical real-world situation where the combination of visual and auditory inputs aids us in reaching a more informed conclusion: imagine rain falling outside of a window. It is often difficult to correctly estimate the density of rain by just looking outside a closed window. It is also not easy to estimate the amount of rain when only listening to it with your eyes closed. It is the holistic audiovisual perspective, perceptually integrating both of our senses, that allows us to best determine whether we should use an umbrella or even stay indoors.

Inspired by the capabilities of the human visual and auditory systems and the possibility of integrating visualization and sonification, this STAR covers academic contributions from both the visualization and the sonification communities that blend sonification and visualization within the context of data exploration and data presentation. We hope this STAR will help both visualization and sonification researchers realize the potential of such combinations and foster future collaborations between two often \revized{disjoint} communities.

\subsection{Sonification Background}
This STAR being published at a visualization venue calls for a brief introduction to the sonification techniques that are part of our data (for a more comprehensive description of the different techniques, see the Sonification Handbook~\cite{hermann_2011_handbook}). \revized{Overall, we can distinguish the five main techniques of \emph{audification}, \emph{parameter mapping sonification}, \emph{model-based sonification}, \emph{auditory icons}, and \emph{earcons}:} 

\revized{\emph{Audification} is a technique to represent typically long sequences of data values (often time-series) by interpreting them as digital audio waveforms and directly playing them back over a loudspeaker. The resulting sound is a translation of the data values into the audible domain in terms of frequency and loudness. An example of audification is the playback of seismographic data with increased playback speed, such that the original low-frequency signal gets pitch-shifted into a range that is audible to humans. A design challenge for audification is the influence of the chosen playback speed on the salience of emerging auditory patterns that should inform a listener about their data.}

\revized{\emph{Parameter mapping sonification} (PMS) is a technique that involves the association of data values with auditory parameters such as pitch, timbre, and loudness. The technique is conceptually closest to many standard visualization techniques as it employs the direct mapping of data values to auditory channels of a carrying sound. If a visualization utilizes a visual mark to represent information by the mark's position or its color, then a parameter mapping sonification utilizes a sound to carry information via its pitch, loudness, or other auditory channels. Similar to a visualization, such a sound is dependent on the mapping function between the data and the auditory channel as well as the nature of the mapping, for example, being linear to linear or linear to exponential. An open challenge in sonification research is the task-dependent and appropriate selection of auditory channels or the perceptual influence of one channel on another. In terms of visualization, this is comparable to the influence of the spatial size of a mark on our perception of its color.}

\revized{The technique of \emph{model-based sonification} (MBS) is inspired by the fact that most of the interactions with our physical environment result in an acoustic response. These acoustic responses, such as the sound of a drum being hit, inform us about the state of the object that we interact with (in the case of a drum, its tuning). Model-based sonification is a general term for sonification techniques that make use of dynamic models describing changes in a system over time. These dynamic models are ``tuned'' by the data that an analyst wants to explore. To listen to their data, a user is required to excite the model with an interaction, such as when a drummer needs to hit their instrument to hear its tuning. Another example is the excitation of a mass-spring model, where the mass and spring parameters are determined by high-dimensional data, defining the sound of the model when excited by a user. The technique is meant to foster exploratory data analysis, as the type and place of interaction strongly influence the acoustic response of the model. In this context, an ongoing research challenge is how to establish an intuitive relation between excitation modes (where and how hard to hit the drum) and typical interactions during exploratory data analysis, such as zooming or filtering of the data.}

\revized{Also inspired by real-world sounds, \emph{auditory icons} are short, distinctive sounds present in everyday life that can be compared to visual icons. This means that there is an inherent association between the auditory icon and the event they represent. A classic example of an auditory icon is the sound of a piece of paper being crumpled and thrown into a bin, representing deleting a file on a computer. A research challenge is the possibility of cultural differences between listeners, as they can lead to confusion or misinterpretation of auditory icons~\cite{jeon_2015_CulturalDifferencesPreference}, similar to the way visual icons can be context-dependent.}

\revized{\emph{Earcons}, comparable to visual symbols, are short, distinctive sounds or melodies that are often used to represent specific events. These sounds are usually synthesized tones or sound patterns and can be described as designed or composed sound symbols. Since there is no inherent association with the real world, the meaning of the earcon needs to be learned before it can be beneficially used. Examples of such earcons are the sounds provided to pilots in the cockpit of an airplane, alerting them about events that require attention.}

\new{While other techniques for the sonification of data exist, these five are the most prominent ones. Within the scope of our STAR, we identified all five of these techniques, even though the vast majority of papers utilize parameter mapping sonification. Some of the papers also combine techniques (for example, parameter mapping sonification and auditory icons, see \autoref{tab:data overview}).}

\noindent
\textbf{A brief history of sonification:}
While the discipline of visualization has a relatively long history~\cite{friendly_2008_BriefHistoryData}, the research field of sonification is younger~\cite{frysinger2005brief}. \new{In 1982, Bly submitted her PhD thesis Sound and computer information presentation~\cite{bly1982sound}, where she suggested methods of encoding information into sound.} \revized{Ten years later, the first International Conference on Auditory Display convened, which is regarded as the birth of the International Community For Auditory Display (ICAD). When its proceedings were published in 1994, the book Auditory Display~\cite{kramer1994auditory} was a reflection on the potential of the newborn field of research that encompasses sonification.}
Early on, Barrass~\cite{barrass1997auditory} presented a task taxonomy for auditory displays, called \textit{TaDa!}, which stands for Tasks and Data. The \textit{TaDa!} taxonomy is especially relevant in the context of this STAR, as it is well-aligned with taxonomies from the visualization literature~\cite{bertin_1983_semiology, brehmer_2013_multi, schulz_2013_design, shneiderman_1996_eyes,yi_2007_towarddeeper}, and it also functions as inspiration for the classification applied later in this STAR. 

The book Ecological Psychoacoustics, edited by Neuhoff in 2004~\cite{neuhoff_ecological_2004}, challenged many psychoacoustical studies (which is the part of psychophysics that involves the scientific study of sound perception, traditionally conducted in controlled laboratory environments). Neuhoff promoted an ecological sound approach to sonification from a holistic perspective, which echoes the aims of the BELIV workshop series established in 2006 at the Advanced Visual Interfaces conference~\cite{beliv2006proc}. Neuhoff's intervention underscores the need to consider real-world contexts for transforming design principles and methodologies for auditory displays from theory to practice. This perspective emphasized integrating ecological factors in sonification to increase effectiveness and deepen the connection between auditory stimuli and real-world experiences.

The \textit{sonification design space map} was introduced in 2007 by deCampo~\cite{deCampo2007toward}, guiding a designer's decision-making process of selecting an appropriate sonification technique for their task. The map creates a two-dimensional space between the number of data properties a designer intends to sonify and the number of data points that are necessary for the different sonification techniques to be employed adequately. Retrospectively, another milestone within the sonification community was the introduction of a now widely accepted definition of sonification as a scientific technique for representing data, presented by Hermann~\cite{hermann2008taxonomy} in 2008. Before the introduction of this definition, it was less clear where to draw the border between artistic and scientific mappings from data to sound (which brings to mind the discussion that data visualization is more than just pretty pictures). As it reflects our understanding of the term sonification, we want to refer to the full definition below:

\textit{``A technique that uses data as input, and generates sound signals (eventually in response to optional additional excitation or triggering) may be called sonification, if and only if
\begin{itemize}
    \item The sound reflects objective properties or relations in the input data.
    \item The transformation is systematic. This means that there is a precise definition provided of how the data (and optional interactions) cause the sound to change.
    \item The sonification is reproducible: given the same data and identical interactions (or triggers) the resulting sound has to be structurally identical.
    \item The system can intentionally be used with different data, and also be used in repetition with the same data.''
\end{itemize}}

The Sonification Handbook~\cite{hermann_2011_handbook}, published in 2011, provided the first general and overarching perspective on the field of sonification, discussing both the theory and practice of sonification. The Handbook is still the most comprehensive collection of sonification work, and therefore, its publication year in 2011 also marks the beginning of the time period considered in our STAR.

Over the years, various design frameworks for sonification have been proposed. The design framework proposed by Barrass~\cite{barrass_2012_aesthetic} emphasizes the fusion of aesthetics and functionality to improve the accessibility and meaningfulness of sonifications for a broader audience. The work of Worrall in 2019~\cite{worrall_sonification_2019} formalized sonification techniques into a framework that also highlighted the challenges and advantages of these sonification techniques, as well as the importance of understanding processes and choices that influence sound representation. The \textit{sonification design canvas}, introduced by Lenzi in 2021~\cite{lenzi_2021_design}, is a contribution to the construction of a more comprehensive design framework, with the aim of integrating all aspects into a cohesive design tool. Despite these efforts, developing a comprehensive protocol that systematically considers end-users at each stage of the design process has yet to be achieved. 

More than 30 years after the beginning of systematic sonification research, we saw a considerable number of theoretical contributions to the field~\cite{kramer1994auditory, kramer_1999_report, vickers_2006_sonification, deCampo2007toward, nees2008encoding, hermann_2011_handbook, grondSingingFunctionExploring2012, supper_2012_lobbying, nees_2019_eight, neuhoff2019sonification, worrall_sonification_2019, lenzi_2021_design}. Explicit work integrating sonification and visualization theory is rare\new{, but has been called for \cite{roberts_2010_UsingAllOur}}. In an attempt to find a common language and, consequently, build a theoretical bridge between the visualization and the sonification communities, Enge et al.~\cite{enge_2023_unified} introduced three theoretical constructs to formally describe audiovisual display idioms. 
They defined the ``auditory mark'' inspired by the visual mark, the ``auditory channel'' inspired by the visual channel, and the ``substrate of sonification'' inspired by the spatial substrate that is used in visualization theory~\cite{bertin_1983_semiology, card_1999_readings}. The definition of \textit{time} as the substrate of sonification allows a description of sonification designs with auditory marks placed in time, with data encoded into their auditory appearance using auditory channels such as pitch or loudness. These definitions allow a high-level discussion and categorization of both the visual and the auditory part of an audiovisual display idiom. The theoretical constructs proved useful, so we adopted the term ``auditory channels'' for classification in this STAR as well.
\subsection{\new{Motivation and related work}}

\new{Seminal visualization texts such as Wilkinson's Grammar of Graphics \cite{wilkinson_2005_grammar} and Spence's textbook \cite{spence_2007_information} made clear statements that their understanding of visualization respectively graphics is not limited to vision but that data can be encoded for other sensory modalities such as sound.
Already at the 6th IEEE Visualization Conference, Minghim and Forrest \cite{minghim_1995_illustrated} postulated areas where sonification can help tackle visualization challenges such as adding complementary or redundant dimensions, natural mapping for time-oriented data, or improved memory of data. 
Some research agendas \cite{thomas_2005_illuminating} suggest multimodal approaches so that one sensory modality can overcome problems that others may have. 
In 1990, Grinstein and colleagues \cite{grinstein_1990_perceptualization, smith_1990_stereophonic} presented an audiovisual interface for the exploration of multivariate data using icons and parameter mapping sonification.  
Also, several works presented at early visualization conferences have integrated sound into visualizations for surfaces, volumes, and fluid dynamics \cite{minghim_1995_illustrated, lodha_1996_listen, rossiter_1996_system, volpe_1997_auralization, frohlich_1999_exploring}. 
The \textit{Data Mountain} interface \cite{robertson_1998_data} augmented its spatial document management space with auditory cues that indicated how many pages were moving. Published as early as 1989, Gaver presented the \textit{Sonic Finder,} which was the auditory user interface used in Apple computers and coined the term ``auditory icon.''}

\new{The auditory perception has an exceptional ability to detect temporal changes and patterns~\cite{guttman_2005_hearing, Rubinstein_1971_crossmodalpatterns}. Also, human hearing is capable of perceiving and distinguishing between several sounds simultaneously, at least between three auditory streams at the same time~\cite{walker_2013_streams}. Another capability of our auditory perception is the possibility to detect and focus on events that spatially occur all around the listener, which can enable an information display to convey peripheral information to a user. Furthermore, the auditory system enables quick reactions when performing certain types of tasks due to the different processing times of the senses~\cite{Jain_2015_responsetime}.}

\new{The combination of audio and visual data representations has shown to be advantageous, particularly for a number of applications.
First, situations where the visual modality is busy with another task, such as monitoring~\cite{iber_2020_handbook, nesbitt_evaluation_2002}, lend themselves beneficial for being complemented with sonification. 
Second, the combination of visual and auditory techniques has been shown to better facilitate learning \cite{Mayer_2014_multimedialearning, Seitz_2006_soundvisuallearning}, since it has the potential to increase working memory capacity and retention of information while also reducing cognitive load.
Third, using sonification also shows benefits for data exploration. Flowers et al.~\cite{flowers_1997_crossmodal} published a seminal article in the sonification community that demonstrated that auditory scatter plots, where data is mapped to onset time and pitch of sounds instead of horizontal and vertical position of visual marks, can offer similar performance as visual representations.}

\new{Research has shown that visual and auditory perception is naturally integrated with each other, which can be observed, for example, with so-called crossmodal correspondences~\cite{Spence_2011_crossmodal,spence_2023_SensoryTranslationAudition}. In our context, crossmodal correspondences describe a phenomenon where we perceive different visual and auditory stimuli as inherently related to each other. Such correspondences could provide an opportunity to use the strengths of the auditory modality to support and enhance visualizations, creating more effective and compelling representations of data~\cite{kasakevich_2007_augmentation}. Rosli and Cabrera~\cite{Rosli_2015_GPM}, for example, have identified the potential of integrated designs to form a more concise, general representation of the data set compared to individual stimuli alone. Rubab et al. recently explored relationships between auditory channels and visual channels and suggested factors that impact their effectiveness~\cite{rubab_2023_ExploringEffectiveRelationshipsc}.}

\new{Overall, the research described above suggests that the integration of the auditory modality has the potential to remedy challenges that exist for visual perception. Challenges, such as simultaneous brightness contrast~\cite{ware_2019_information} or the Mach band phenomenon~\cite{lotto_1999_mach}, impact perception of visual representations~\cite{schloss_2018_mapping,zimnicki_2023_effects}. It has been demonstrated that various auditory channels can be successfully linked and related to visual channels~\cite{ferguson_2018_investigating,ward_2006_soundcolour, collier_2004_musical}, which could offer a way of substituting visual channels with auditory channels to address these challenges.}

\subsection{\revized{Related Surveys}}
\label{Sec: Related Work} 

In general, systematic state-of-the-art reports are less established in the sonification community. A rare exception is the ``\textit{systematic review of mapping strategies for the sonification of physical quantities}'' by Dubus and Bresin~\cite{dubusSystematicReviewMapping2013}, which, however, just covers sonification-only contributions. Much earlier, in 2001, Walker and Lane~\cite{walker2001sonification} provided a website enabling researchers to search for sonification mappings that have been used in scientifically evaluated designs. Unfortunately, the website is no longer available.
Another more timely and exhaustive exploration of sonification literature emerges in Andreopoulou and Goudarzi's 2021 publication~\cite{andreopoulou_2021_sonification}. The authors reviewed 456 papers from the International Conference on Auditory Display proceedings. This incisive analysis exposes compelling trends, ranging from the sonification domains to the diverse publication venues. The report reveals linguistic trends and explores the balance between research and artistic contributions. In addition, it illuminates the landscape of tools, methodologies, and evaluation practices that have led to sonification's multifaceted evolution.
Marking the beginning of an important sociocultural reflection within the sonification community, in 2017, Andreopoulou and Goudarzi~\cite{andreopoulou_2017_reflections} also studied the ``\textit{representation of female researchers and artists in the conferences of the International Community for Auditory Display (ICAD)}''. Their findings showed that only about $18\%$ of ICAD papers were co-authored by women, with stagnant numbers between the years 1994 and 2016.

With respect to combinations of visualization and sonification, Caiola et al.~\cite{caiola_2022_AudiovisualSonificationsDesign} recently presented an analysis of visual and auditory channels commonly used in audiovisual display idioms. Their survey includes combined idioms that map data attributes redundantly to both a visual channel (such as position) and an auditory channel (such as pitch). Analyzed works stem from the Sonification Archive (described below) and a Google keyword search using sonification-related terms exclusively. 
The \href{https://sonification.design/}{\textit{Sonification Archive}}, widely known in the sonification community, is a curated collection of sonification designs, often related to other modes of representation, such as visualization. The Sonification Archive holds both artistic and academic contributions, as well as designs from data journalism. 

Searching through the visualization literature, we were not able to find any STAR or survey focused on the integration of sonification and visualization. We explored the survey of surveys~\cite{mcnabb_2017_SurveySurveysSoS} but could not identify any related contributions. Therefore, to the best of our knowledge, this is the first systematic STAR dedicated to academic contributions in the intersection of sonification and visualization for data exploration and presentation.

\subsection{How to Use This Survey}

With this STAR, we intend to provide an overview of an emerging research field, as well as connect two mostly disjoint research communities. We hope to reach researchers from both communities, inspiring them to intertwine sonification and visualization in their future research. We see several ways of using this STAR:
\begin{itemize}
    \item using it as an overview, intended for researchers who seek a summary of the field.
    \item using it to find research opportunities and existing gaps in the field.
    \item using our supplemental material to study the existing meta-data in more detail, such as identifying authors from the respective other field for potential collaboration. Furthermore, we provide a public \href{https://www.zotero.org/groups/integrationsonificationvisualization/items}{\textit{Zotero} library}, holding all relevant publication metadata, our tags, and all open access PDFs. 
\end{itemize}

This STAR will be structured as follows:
\autoref{sec: Method} describes the methodology used to search and filter the literature identified as potentially relevant. In \autoref{sec: categorization and results}, we describe our classification system and use it to discuss the survey literature. In \autoref{Sec: SurveyDataAnalyes}, we apply a meta-perspective on the survey data, describing correlations between individual tags, as well as the co-author network of the field. In \autoref{sec: adjacent} we introduce the three adjacent topics of accessibility, monitoring, and arts, which are related to our STAR, but were not studied systematically. Finally, in \autoref{sec: discussion}, we offer a concluding discussion focusing on future work.

\section{Method} 
\label{sec: Method}
In this section, we discuss our inclusion and exclusion criteria and the methods we used to search for the relevant literature. We used a five-stage pipeline to construct a corpus of research that is at the intersection of visualization and sonification for data exploration and presentation (see \autoref{fig:literature filter process}).

\subsection{Scope of the Surveyed Literature}
Sonification and visualization share the aim of making data interpretable to their users and observers. With this shared goal, combinations of the two can be designed for numerous possible applications and contexts. Our research interest in this STAR is the combination of sonification and visualization in the context of data analysis, covering both data exploration and presentation. A work that is relevant to our STAR must include both visualization and sonification of data. Therefore, a sonification with a visual interface that does not represent data is not enough to be considered relevant, and neither is a visualization with sounds that do not represent data. The work must be an academic paper published between the years 2011 and 2023 and must be peer-reviewed to be considered in our STAR.

Thinking more broadly about the combination of sonification and visualization, three additional areas of application come to mind: (1) accessibility, (2) real-time monitoring, and (3) arts. 
All three areas are vast, and a detailed classification of them is beyond the scope of our STAR. However, we find them relevant and inspirational for our field.
Thus, we provide a brief introduction to the fields of accessibility, monitoring, and artistic contributions in \autoref{sec: adjacent}. In the context of accessibility, sonification can be used to support the collaboration between blind and sighted users by mapping data to both an auditory and a visual display. In the same manner, such a design could support the collaboration between deaf individuals and individuals without hearing impairment.
Nevertheless, our research interest is the combination of sonification and visualization for the integrated analysis of data. Therefore, in our STAR, we consider only designs intended to be used with both the visual and the auditory senses fully available to a user. The application of real-time monitoring, such as medical monitoring, critical infrastructure monitoring, alarms, or real-time feedback on body movement, is vast and distinct from the purpose of data exploration. Especially with respect to sonification and auditory display, the field is well-researched (e.g.,~\cite{kanev_sonification_2019, schaffert_review_2019, van_rheden_sonification_2020, wang_interactive_2017, hildebrandt_server_2015}), and we will not cover such designs in this STAR. Artistic contributions have the potential to be highly inspirational for our field but require a different search method and, most likely, a different system of classification. Again, we decided not to systematically cover artistic contributions in this STAR but to provide a subsection discussing a list of representative works that may serve as a starting point for future research interests.

\subsection{Search Strategy and Filtering}
\label{sec: Search strategy and filtering}

\begin{figure}
    \centering
    \includegraphics[width=\linewidth]{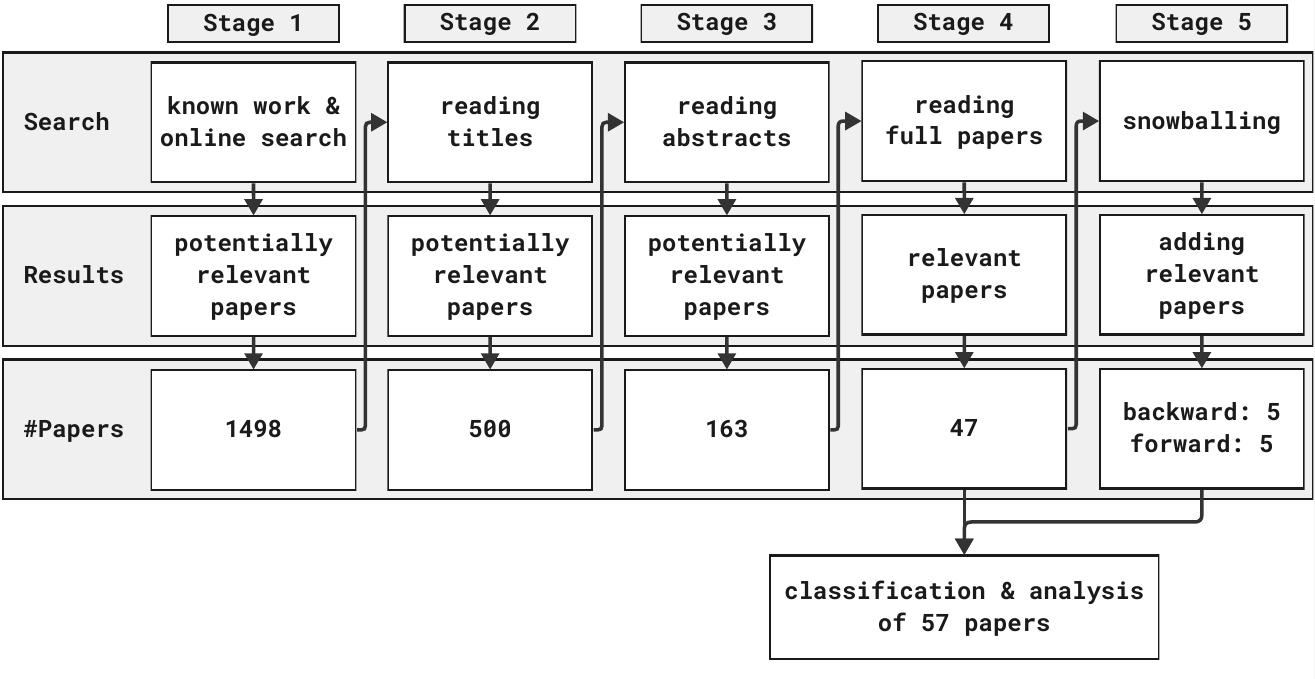}
    \caption{The literature search was conducted in five stages, ranging from an online keyword search to using the snowballing technique on the papers identified as relevant. \new{Figure inspired by \cite{Yeshchenko_2022_STAR}.}}
    \label{fig:literature filter process}
\end{figure}

We base our corpus of literature on (1) publications that the authors have already been aware of from their previous work in this field and (2) an extensive online literature search. The online search was a keyword-based search in the digital libraries of IEEE Xplore, ACM Digital Library, and Springer Link, which include work published at IEEE VIS, CHI, and other VGTC- and SIGCHI-sponsored venues. Furthermore, we searched the digital libraries of Eurographics, ICAD, ISon, Organised Sound, and the Sound and Music Computing Community. 
\autoref{fig:literature filter process} provides an overview of the different stages we used to systematically filter our database for relevant publications. To keep track of the progress, we used a Google sheet document. The final database, including all papers and all tags by the authors, is provided in the supplemental material as a CSV file. 

\noindent
\textbf{Stage 1 --} The search query we used in stage 1 was the following: \textit{("Visualization" OR "Visualisation" OR "Visual Analytics") AND ("Sonification" OR "Auditory Display") AND NOT (centrifug* OR lys* OR homogeniz*).} It consists of an AND combination of a visualization term and a sonification term combined with the exclusion of specific word beginnings. \revized{While the terms ``visualization'' and ``visual analytics'' are well known, our audio-related search terms are likely to be less well-known overall. Nevertheless, we decided not to include search terms such as music, sound, or tone in our online search. Searching for publications with such broad terminology would have resulted in too many papers to be scanned. Also, we argue that academic publications interested in the integration of visualization and sonification are likely using the appropriate terminology.}
The reason to exclude papers that hold words starting with \textit{centrifug*}, \textit{lys*}, and \textit{homogeniz*} is the fact that sonification, also called sonication, is a term also used in biology to describe a process where sound is used to agitate particles in a sample. As papers using sonification in this context are not relevant to this STAR, we identified terms, including the three mentioned above, that are often associated with this meaning of sonification.
The search in the Springer Link database was \revized{especially sensitive} to the exclusion of these terms. As there could be false negatives using the exclusion of the three terms, we manually reviewed the paper titles that were excluded due to this strategy and restored five potentially relevant papers.
This online search, combined with papers we were already aware of from our previous work in the field, resulted in a database holding 1498 papers. We used the literature management software \href{https://www.zotero.org/}{Zotero} to download the respective papers and to make them available to all co-authors. 

\noindent
\textbf{Stage 2 --} In the second phase of our literature search, each paper title was read by two of the authors and classified into \textit{potentially relevant} or \textit{irrelevant}. We agreed to use an inclusive mindset for this stage, i.e., we tagged vague titles mostly as potentially relevant so as not to overlook many papers at this early stage. For papers that were tagged differently by two people, the two people came to an agreement, or the paper was taken to the next stage. Stage 2 resulted in a database holding 500 papers. 

\noindent
\textbf{Stage 3 --} In the third phase of our literature search, each abstract was read by two of the authors and classified into \textit{potentially relevant} or \textit{irrelevant}. Again, we used an inclusive mindset and tagged vague abstracts as potentially relevant. For papers that were tagged differently by two people, the two people came to an agreement, or the paper was taken to the next stage. Stage 3 resulted in a database holding 163 papers. 

\noindent
\textbf{Stage 4 --} In the fourth phase of our literature search, each paper was read by one of the authors and classified as \textit{relevant} or \textit{irrelevant}. As the authors had a solid common understanding of the inclusion and exclusion criteria by this stage, and as a person had the full paper information available to make a decision, one person decided on the relevance at this stage. 
The papers identified as relevant in stage 4 were classified using the tags explained in detail in \autoref{tab:tag-definitions}. Each paper was first classified by one of the co-authors, and their tags were later verified by a second co-author. Whenever two co-authors initially disagreed on a specific tag, they came to an agreement, or the first author made a decision. Stage 4 resulted in a database holding 47 papers. 

\noindent
\textbf{Stage 5 --} Furthermore, we extended the corpus of relevant papers by snowballing~\cite{wohlin_2014_GuidelinesSnowballingSystematic}, checking all incoming and outgoing references of the articles matching our inclusion criteria. Snowballing was done by one of the co-authors with an exclusive mindset towards the paper titles, meaning that a vague title was not considered relevant. Stage 5 resulted in our database holding 57 papers overall, adding ten papers to the prior stage. 
During the final two stages, whenever we identified an audiovisual idiom that was published in more than one paper, we retained the most extensive version in our STAR. We identified two such cases where a design was previously published in a short paper but later expanded~\cite{yang_2018_InteractiveModeExplorer, malikova_2019_VisualauditoryVolumeRendering}. \new{It is notable that a considerable number of papers that are part of our final scope were published at the International Conference on Auditory Display (ICAD). This is not surprising as this venue is the most recognized venue to publish sonification work holding the largest single corpus of work in the field~\cite{bearman_2012_WHOSONIFYINGDATAb,andreopoulou_2021_sonification,Gross-Vogt_2023_EffectiveSonification}. It is also not surprising that it is a sonification venue that is prominent, as there is a structural imbalance between the domains. Many sonification designs, in general, also include some sort of visualization in their design, but that is not true the other way around.}

\begin{table*}[]
\centering
\resizebox{\textwidth}{!}{%
\begin{tabular}{ccccccccccccccccccc}
\hline
  \multirow{2}{*}{\textbf{Paper}} & \multirow{2}{*}{\textbf{Theme}} & \multirow{2}{*}{\textbf{Demo Link}} & \multicolumn{3}{c}{\multirow{2}{*}{\textbf{Purpose}}} & \multicolumn{3}{c}{\multirow{2}{*}{\textbf{Redundancy}}} & \multirow{2}{*}{\textbf{Visualization Idiom}} & \multicolumn{5}{c}{\multirow{2}{*}{\textbf{Sonification Technique}}} & \multicolumn{4}{c}{\multirow{2}{*}{\textbf{Goal}}} \\ \\ \cline{4-9} \cline{11-19} 
  \multicolumn{1}{c}{}          & \multicolumn{1}{c}{}           & \multicolumn{1}{c}{}                & \rotatebox[origin=c]{90}{Exploration} & \rotatebox[origin=c]{90}{Presentation} & \multicolumn{1}{c|}{\rotatebox[origin=c]{90}{Both}} &
  {\rotatebox[origin=c]{90}{Redundant}} &
  {\rotatebox[origin=c]{90}{Mixed}} &
  {\rotatebox[origin=c]{90}{Complementary}} &
  \multicolumn{1}{c}{} &
  {\rotatebox[origin=c]{90}{PMS}} &
  {\rotatebox[origin=c]{90}{MBS}} &
  {\rotatebox[origin=c]{90}{Audification}} &
  {\rotatebox[origin=c]{90}{Auditory Icons}} &
  \multicolumn{1}{c|}{\rotatebox[origin=c]{90}{Earcons}} &
  {\rotatebox[origin=c]{90}{Education}} &
  {\rotatebox[origin=c]{90}{Data Analysis}} &
  {\rotatebox[origin=c]{90}{Research}} &
  \multicolumn{1}{c}{\rotatebox[origin=c]{90}{~~Public Engagement~~}}
  \\ \hline
  
\cite{elmquist_2021_OpenspaceSonificationComplementing} & Astronomy                       & \href{https://vimeo.com/528822742}{[<—>]}  &                                     &                                      & $\bullet$                         &                                   & $\bullet$                             &                                       & volume rendering                 & $\bullet$                       &                             &                                      &                                        &                              &                                   &                                       &                                  & $\bullet$                                      \\ \hline
\cite{garciariber_2019_SonifigrapherSonifiedLight} & Astronomy                       & \href{https://archive.org/details/SonifigrapherMacOSX}{[<—>]}  &                                     & $\bullet$                                    &                           & $\bullet$                                 &                               &                                       & line chart, scatter plot         & $\bullet$                       &                             &                                      &                                        &                              & $\bullet$                                 &                                       &                                  &                                        \\ \hline
\cite{harrison_2022_AudioUniverseTour} & Astronomy                       & \href{https://www.youtube.com/watch?v=5HS3tRl2Ens}{[<—>]}  &                                     & $\bullet$                                    &                           & $\bullet$                                 &                               &                                       & point cloud                      & $\bullet$                       &                             &                                      &                                        &                              &                                   &                                       &                                  & $\bullet$                                      \\ \hline
\cite{huppenkothen_2023_SonifiedHertzsprungRussellDiagram} & Astronomy                       & \href{https://starsounder.space/}{[<—>]} &                                     &                                      & $\bullet$                         &                                   &                               & $\bullet$                                     & scatter plot                     & $\bullet$                       &                             & $\bullet$                                    &                                        &                              & $\bullet$                                 &                                       &                                  &                                        \\ \hline
\cite{riber_2018_PlanethesizerApproachingExoplanet} & Astronomy                       & \href{https://archive.org/details/PlanethesizerWindows}{[<—>]} &                                     &                                      & $\bullet$                         & $\bullet$                                 &                               &                                       & volume rendering                 & $\bullet$                       &                             &                                      &                                        &                              &                                   & $\bullet$                                     &                                  &                                        \\ \hline
\cite{russo_2022_5000ExoplanetsListen} & Astronomy                       & \href{https://exoplanets.nasa.gov/resources/2321/5000-exoplanets-listen-to-the-sounds-of-discovery-360-video/}{[<—>]} &                                     & $\bullet$                                    &                           &                                   & $\bullet$                             &                                       & 3D map                           & $\bullet$                       &                             &                                      &                                        &                              &                                   &                                       &                                  & $\bullet$                                      \\ \hline
\cite{traver_2023_HarmonicesSolarisSonification} & Astronomy                       &   &                                     & $\bullet$                                    &                           &                                   &                               & $\bullet$                                     & 3D scatter plot                  & $\bullet$                       &                             &                                      &                                        &                              &                                   &                                       &                                  & $\bullet$                                      \\ \hline
\cite{bru_2023_LineHarpImportanceDriven} & Domain Agnostic Display    &  & $\bullet$                                   &                                      &                           & $\bullet$                                 &                               &                                       & line chart, parallel coordinates & $\bullet$                       &                             &                                      &                                        &                              &                                   & $\bullet$                                     &                                  &                                        \\ \hline
\cite{cantrell_2021_HighchartsSonificationStudio} & Domain Agnostic Display    & \href{https://sonification.highcharts.com/}{[<—>]} &                                     &                                      & $\bullet$                         & $\bullet$                                 &                               &                                       & line chart                       & $\bullet$                       &                             &                                      &                                        &                              & $\bullet$                                 & $\bullet$                                     &                                  & $\bullet$                                      \\ \hline
\cite{delavega_2022_SonoUnoWebInnovative} & Domain Agnostic Display    & \href{https://www.sonouno.org.ar/}{[<—>]}  &                                     &                                      & $\bullet$                         & $\bullet$                                 &                               &                                       & line chart                       & $\bullet$                       &                             &                                      &                                        &                              &                                   & $\bullet$                                     &                                  &                                        \\ \hline
\cite{du_2018_ExploringRoleSound} & Domain Agnostic Display    &  &                                     &                                      & $\bullet$                         & $\bullet$                                 &                               &                                       & convex hull, other, bar chart    & $\bullet$                       &                             &                                      &                                        &                              &                                   & $\bullet$                                     &                                  & $\bullet$                                      \\ \hline
\cite{enge_2022_MultimodalExploratoryData} & Domain Agnostic Display    & \href{https://phaidra.fhstp.ac.at/detail/o:4831}{[<—>]} & $\bullet$                                   &                                      &                           & $\bullet$                                 & $\bullet$                             & $\bullet$                                     & scatter plot                     & $\bullet$                       &                             &                                      &                                        &                              &                                   & $\bullet$                                     &                                  &                                        \\ \hline
\cite{ferguson_2012_NavigationInteractiveSonifications} & Domain Agnostic Display    &  &                                     &                                      & $\bullet$                         & $\bullet$                                 &                               &                                       & line chart                       & $\bullet$                       &                             &                                      &                                        &                              &                                   & $\bullet$                                     &                                  &                                        \\ \hline
\cite{groppe_2021_SoundDatabasesSonification} & Domain Agnostic Display    & \href{https://www.ifis.uni-luebeck.de/~groppe/soundofdatabases}{[<—>]} & $\bullet$                                   &                                      &                           &                                   &                               & $\bullet$                                     & network                          & $\bullet$                       &                             &                                      &                                        &                              & $\bullet$                                 &                                       &                                  & $\bullet$                                      \\ \hline
\cite{kondak_2017_WebSonificationSandbox} & Domain Agnostic Display    & \href{https://youtu.be/BhL3J5hcwNE?t=7650}{[<—>]} &                                     & $\bullet$                                    &                           &                                   & $\bullet$                             &                                       & line chart                       & $\bullet$                       &                             &                                      &                                        &                              &                                   &                                       &                                  & $\bullet$                                      \\ \hline
\cite{lindetorp_2021_SonificationEveryoneEverywhere} & Domain Agnostic Display    & \href{https://hanslindetorp.github.io/SonificationToolkit/}{[<—>]} &                                     &                                      & $\bullet$                         & $\bullet$                                 &                               &                                       & line chart                       & $\bullet$                       &                             &                                      &                                        &                              & $\bullet$                                 & $\bullet$                                     &                                  & $\bullet$                                      \\ \hline
\cite{malikova_2019_VisualauditoryVolumeRendering} & Domain Agnostic Display    & \href{https://vimeo.com/323545930}{[<—>]*} & $\bullet$                                   &                                      &                           & $\bullet$                                 &                               &                                       & heatmap                          & $\bullet$                       &                             &                                      &                                        &                              &                                   & $\bullet$                                     &                                  &                                        \\ \hline
\cite{peng_2023_SirenCreativeExtensible} & Domain Agnostic Display    & \href{https://kizjkre.github.io/siren/}{[<—>]} &                                     & $\bullet$                                    &                           & $\bullet$                                 &                               &                                       & line chart                       & $\bullet$                       &                             &                                      &                                        &                              & $\bullet$                                 &                                       &                                  &                                        \\ \hline
\cite{phillips_2019_SonificationWorkstation} & Domain Agnostic Display    &  &                                     & $\bullet$                                    &                           & $\bullet$                                 &                               &                                       & line chart                       & $\bullet$                       &                             &                                      &                                        &                              &                                   & $\bullet$                                     &                                  &                                        \\ \hline
\cite{ronnberg_2016_InteractiveSonificationVisual} & Domain Agnostic Display    & \href{https://vimeo.com/247770770}{[<—>]} & $\bullet$                                   &                                      &                           & $\bullet$                                 &                               &                                       & scatter plot, parallel coordinates                     & $\bullet$                       &                             &                                      &                                        &                              &                                   & $\bullet$                                     &                                  &                                        \\ \hline
\cite{yang_2018_InteractiveModeExplorer} & Domain Agnostic Display    & \href{https://pub.uni-bielefeld.de/record/2920473}{[<—>]} & $\bullet$                                   &                                      &                           &                                   &                               & $\bullet$                                     & scatter plot                     &                         & $\bullet$                           &                                      &                                        &                              &                                   & $\bullet$                                     &                                  &                                        \\ \hline
\cite{fitzpatrick_2018_StreamSegregationUtilizing} & Domain Agnostic Display        &  & $\bullet$                                   &                                      &                           &                                   & $\bullet$                             &                                       & line chart                       & $\bullet$                       &                             &                                      &                                        &                              &                                   & $\bullet$                                     &                                  &                                        \\ \hline
\cite{ballora_2015_TwoExamplesSonification} & Earth Science                   &  &                                     & $\bullet$                                    &                           &                                   & $\bullet$                             &                                       & heatmap                          & $\bullet$                       &                             &                                      &                                        &                              &                                   &                                       &                                  & $\bullet$                                      \\ \hline
\cite{bearman_2011_UsingSoundRepresent} & Earth Science                   & \href{https://vimeo.com/17029358}{[<—>]} & $\bullet$                                   &                                      &                           & $\bullet$                                 &                               &                                       & map                              & $\bullet$                       &                             &                                      &                                        &                              &                                   & $\bullet$                                     &                                  &                                        \\ \hline
\cite{bearman_2012_UsingSoundRepresent} & Earth Science                   & \href{https://www.nickbearman.me.uk/academic/bearman_fisher_2011/index.htm}{[<—>]} & $\bullet$                                   &                                      &                           & $\bullet$                                 &                               &                                       & map                              & $\bullet$                       &                             &                                      &                                        &                              &                                   & $\bullet$                                     &                                  &                                        \\ \hline
\cite{gune_2018_GraphicallyHearingEnhancing} & Earth Science                   &  &                                     &                                      & $\bullet$                         & $\bullet$                                 &                               &                                       & map                              & $\bullet$                       &                             &                                      &                                        &                              &                                   &                                       & $\bullet$                                &                                        \\ \hline
\cite{han_2022_FutureRedVisualizing} & Earth Science                   &  &                                     & $\bullet$                                    &                           & $\bullet$                                 &                               &                                       & geographic scatter plot          & $\bullet$                       &                             &                                      &                                        &                              &                                   &                                       &                                  & $\bullet$                                      \\ \hline
\cite{holtzman_2014_SeismicSoundLab} & Earth Science                   &  &                                     & $\bullet$                                    &                           & $\bullet$                                 &                               &                                       & heatmap                          &                         &                             & $\bullet$                                    &                                        &                              & $\bullet$                                 &                                       &                                  &                                        \\ \hline
\cite{matsubara_2016_CollaborativeStudyInteractive} & Earth Science                   &  & $\bullet$                                   &                                      &                           &                                   &                               & $\bullet$                                     & map                              & $\bullet$                       &                             & $\bullet$                                    &                                        &                              &                                   & $\bullet$                                     &                                  &                                        \\ \hline
\cite{ness_2012_Sonophenology} & Earth Science                   & \href{https://www.youtube.com/watch?v=829r3y01XLk)}{[<—>]} & $\bullet$                                   &                                      &                           &                                   &                               & $\bullet$                                     & map                              & $\bullet$                       &                             &                                      &                                        &                              & $\bullet$                                 &                                       &                                  &                                        \\ \hline
\cite{pate_2022_CombiningAudioVisual} & Earth Science                   & \href{https://parthurp.github.io/homepage/SpatialSeismicSoundscapes_article2021.html}{[<—>]} &                                     &                                      & $\bullet$                         &                                   & $\bullet$                             &                                       & dot map                          &                         &                             & $\bullet$                                    &                                        &                              &                                   &                                       & $\bullet$                                &                                        \\ \hline
\cite{svoronos-kanavas_2022_ExploratoryUseAudiovisual} & Earth Science                   & \href{https://vimeo.com/698105264}{[<—>]} &                                     & $\bullet$                                    &                           & $\bullet$                                 &                               &                                       & fluid-like simulation            & $\bullet$                       &                             &                                      &                                        &                              &                                   &                                       &                                  & $\bullet$                                      \\ \hline
\cite{winters_2015_SonificationTohokuEarthquake} & Earth Science                   & \href{https://www.youtube.com/watch?v=3PJxUPvz9Oo}{[<—>]} &                                     & $\bullet$                                    &                           & $\bullet$                                 &                               &                                       & line chart                       &                         &                             & $\bullet$                                    &                                        &                              &                                   &                                       &                                  & $\bullet$                                      \\ \hline
\cite{gionfrida_2016_TripleToneSonification} & Medicine and Health             &  & $\bullet$                                   &                                      &                           &                                   &                               & $\bullet$                                     & slicing, volume rendering        & $\bullet$                       &                             &                                      &                                        &                              &                                   & $\bullet$                                     &                                  &                                        \\ \hline
\cite{gomez_2011_DATASONIFICATIONAPPROACH} & Medicine and Health             &  & $\bullet$                                   &                                      &                           & $\bullet$                                 &                               &                                       & slicing                          & $\bullet$                       &                             &                                      &                                        &                              &                                   & $\bullet$                                     &                                  &                                        \\ \hline
\cite{lemmon_2023_MappingEmergencyDesigning} & Medicine and Health             & \href{https://ericlemmon.net/ison2022-demo-video/}{[<—>]} & $\bullet$                                   &                                      &                           &                                   & $\bullet$                             &                                       & map                              & $\bullet$                       &                             &                                      &                                        &                              &                                   & $\bullet$                                     &                                  &                                        \\ \hline
\cite{macdonald_2018_DataDrivenSonificationCFD} & Medicine and Health             & \href{https://www.youtube.com/watch?v=UmDvnPjnpV4}{[<—>]} & $\bullet$                                   &                                      &                           &                                   &                               & $\bullet$                                     & volume rendering                 & $\bullet$                       &                             &                                      &                                        &                              &                                   & $\bullet$                                     &                                  &                                        \\ \hline
\cite{roginska_2013_ExploringSonificationAugmenting} & Medicine and Health             &  & $\bullet$                                   &                                      &                           & $\bullet$                                 &                               &                                       & slicing                          & $\bullet$                       &                             &                                      &                                        &                              &                                   & $\bullet$                                     &                                  &                                        \\ \hline
\cite{temor_2021_PerceptuallymotivatedSonificationSpatiotemporallydynamic} & Medicine and Health             & \href{https://www.youtube.com/playlist?list=PLRbQXqE-XKzdDOqxnXI4yfo3r_cIcrjNl}{[<—>]} & $\bullet$                                   &                                      &                           &                                   & $\bullet$                             &                                       & volume rendering                 & $\bullet$                       &                             &                                      &                                        &                              &                                   & $\bullet$                                     &                                  &                                        \\ \hline
\cite{arbon_2018_SonifyingStochasticWalks} & Molecular Science               & \href{https://vimeo.com/255391814}{[<—>]} & $\bullet$                                   &                                      &                           &                                   &                               & $\bullet$                                     & 3D molecule rendering            & $\bullet$                       &                             &                                      &                                        &                              &                                   & $\bullet$                                     &                                  &                                        \\ \hline
\cite{ballweg_2016_InteractiveSonificationStructural} & Molecular Science               &  & $\bullet$                                   &                                      &                           &                                   &                               & $\bullet$                                     & 3D molecule rendering            & $\bullet$                       &                             &                                      & $\bullet$                                      &                              &                                   & $\bullet$                                     &                                  &                                        \\ \hline
\cite{bouchara_2020_ImmersiveSonificationProtein} & Molecular Science               &  & $\bullet$                                   &                                      &                           & $\bullet$                                 &                               &                                       & volume rendering                 & $\bullet$                       &                             &                                      &                                        &                              &                                   & $\bullet$                                     &                                  &                                        \\ \hline
\cite{rau_2015_EnhancingVisualizationMolecular} & Molecular Science               &  & $\bullet$                                   &                                      &                           &                                   &                               & $\bullet$                                     & volume rendering                 & $\bullet$                       &                             &                                      & $\bullet$                                      &                              &                                   & $\bullet$                                     &                                  &                                        \\ \hline
\cite{lyu_2021_AIiveInteractiveVisualization} & Others        & \href{https://www.zhuoyuelyu.com/aiive}{[<—>]} & $\bullet$                                   &                                      &                           &                                   &                               & $\bullet$                                     & 3D network                       & $\bullet$                       &                             &                                      &                                        &                              &                                   &                                       &                                  & $\bullet$                                      \\ \hline
\cite{north_2016_UnderstandingGitHistory} & Others        & \href{https://cse.unl.edu/~myra/artifacts/GitVS/vm/}{[<—>]} & $\bullet$                                   &                                      &                           &                                   & $\bullet$                             &                                       & gantt chart                      & $\bullet$                       &                             &                                      &                                        & $\bullet$                            &                                   & $\bullet$                                     &                                  &                                        \\ \hline
\cite{chabot_2017_ImmersiveVirtualEnvironment} & Others                 &  &                                     & $\bullet$                                    &                           &                                   & $\bullet$                             &                                       & bar chart                        & $\bullet$                       &                             &                                      &                                        &                              &                                   & $\bullet$                                     &                                  &                                        \\ \hline
\cite{macas_2018_ConsumptionRhythmMultimodal} & Others                 & \href{https://vimeo.com/270078256}{[<—>]} &                                     & $\bullet$                                    &                           &                                   & $\bullet$                             &                                       & individual circular design       & $\bullet$                       &                             &                                      &                                        &                              &                                   &                                       &                                  & $\bullet$                                      \\ \hline
\cite{alonso-arevalo_2012_CurveShapeCurvature} & Others       & \href{https://www.youtube.com/@SATINproject/videos}{[<—>]} & $\bullet$                                   &                                      &                           & $\bullet$                                 &                               &                                       & volume rendering                 & $\bullet$                       &                             &                                      &                                        &                              &                                   & $\bullet$                                     &                                  &                                        \\ \hline
\cite{herrmann_2020_VisualizingSonifyingHow} & Others            & \href{https://vincentherrmann.github.io/blog/immersions/}{[<—>]} & $\bullet$                                   &                                      &                           & $\bullet$                                 &                               &                                       & network                          & $\bullet$                       &                             &                                      &                                        &                              &                                   &                                       & $\bullet$                                &                                        \\ \hline
\cite{papachristodoulou_2015_AugmentingNavigationComplex} & Others            &  & $\bullet$                                   &                                      &                           &                                   &                               & $\bullet$                                     & 3D network                       & $\bullet$                       &                             &                                      & $\bullet$                                      &                              &                                   & $\bullet$                                     &                                  &                                        \\ \hline
\cite{papachristodoulou_2014_SonificationLargeDatasets} & Others            &  & $\bullet$                                   &                                      &                           &                                   &                               & $\bullet$                                     & network                          & $\bullet$                       &                             &                                      &                                        &                              &                                   & $\bullet$                                     &                                  &                                        \\ \hline
\cite{hildebrandt_2016_CombiningSonificationVisualization} & Others &  & $\bullet$                                   &                                      &                           &                                   & $\bullet$                             &                                       & dotted chart visualization       & $\bullet$                       &                             &                                      &                                        & $\bullet$                            &                                   & $\bullet$                                     &                                  &                                        \\ \hline
\cite{joliat_2013_SpatializedAnonymousAudio} & Others             &  &                                     &                                      & $\bullet$                         &                                   &                               & $\bullet$                                     & 3D scatter plot                  & $\bullet$                       &                             &                                      &                                        &                              &                                   & $\bullet$                                     &                                  &                                        \\ \hline
\cite{kariyado_2021_AuralizationThreeDimensionalCellular} & Others & \href{https://www.youtube.com/watch?v=eFQi3qFxAp8)}{[<—>]} &                                     & $\bullet$                                    &                           & $\bullet$                                 &                               &                                       & 3D point cloud                   & $\bullet$                       &                             &                                      &                                        &                              &                                   &                                       &                                  & $\bullet$                                      \\ \hline
\cite{ronnberg_2021_SonificationConveyingData} & Others                  & \href{https://vimeo.com/397235072}{[<—>]} &                                     & $\bullet$                                    &                           &                                   &                               & $\bullet$                                     & bar chart                        & $\bullet$                       &                             &                                      &                                        &                              &                                   &                                       &                                  & $\bullet$                                      \\ \hline
\cite{adhitya_2011_SonifiedUrbanMasterplan} & Others          &  &                                     & $\bullet$                                    &                           &                                   & $\bullet$                             &                                       & map                              & $\bullet$                       &                             &                                      &                                        &                              &                                   & $\bullet$                                     &                                  &                                        \\ \hline
\cite{berger_2019_CombiningVRVisualization} & Others          &  & $\bullet$                                   &                                      &                           & $\bullet$                                 &                               &                                       & point grid                       & $\bullet$                       &                             &                                      &                                        &                              &                                   & $\bullet$                                     &                                  &                                    \\ \hline 
\end{tabular}%
}
\caption{A table showing all 57 entries in our database, sorted by their thematic cluster. In the columns for the used sonification technique, ``PMS'' stands for parameter mapping sonification, and ``MBS'' stands for model-based sonification. Where available, the demo links point to the supplemental material of the papers (last accessed on $22^{nd}$ of December 2023). *There are two additional demo videos associated with this paper: \href{https://vimeo.com/323547646}{[<—>]} and \href{https://vimeo.com/323547659}{[<—>]}; all three demos require the password: icad2019.}
\label{tab:data overview}
\end{table*}

\section{Categorization and Results}
\label{sec: categorization and results}

In this section, we will discuss the relevant literature in detail from the perspectives of our classification. The systematic integration of sonification and visualization is a wide and diverse research field that is difficult to classify using only a handful of categories. Therefore, we decided to apply an extensive list of tags to the literature to be able to present diverse perspectives on the field, mostly concerning basic research, i.e.,~the basic principles that distinguish the designs/idioms from each other. 

To give readers an initial thematic overview of the field, we will start by briefly describing each of the 57 selected papers in \autoref{sec: thematic corpus overview}. We will do so by clustering the literature to the following topics: astronomy, medicine and health, molecular science, earth science, domain agnostic data displays, and other topics.
We will then continue with a discussion on the purpose that an idiom can be designed for (\autoref{sec: Purpose}), followed by an analysis of idiom design possibilities (\autoref{sec: Audiovisual Idiom design}). We will study several technical perspectives that were tagged individually for the sonification and the visualization aspects of each paper: the reading levels (\autoref{sec: Reading Levels}) suggested by Bertin~\cite{bertin_1983_semiology}, the search levels (\autoref{sec: Search Levels}) suggested by Munzner~\cite{munzner_visualization_2015}, as well as the dataset types~\cite{munzner_visualization_2015} including the levels of measurement of the displayed data (\autoref{sec: dataset types and level of measurement}). We will then review different levels of mapping redundancy (\autoref{sec: Level of Redundancy}), different evaluation methods (\autoref{sec: Evaluation Approaches}), different target platforms, and various possibilities of interacting with audiovisual display idioms (\autoref{sec: Target Plattforms and Interactivity}). Finally, we study the diverse user groups and the possible goals of designers (\autoref{sec: Users and Goals}).

All the above categories and subcategories are concisely presented in two tables.
\autoref{tab:data overview} shows an overview of all papers and a subset of their most relevant tags.
\autoref{tab:tag-definitions} provides a detailed description for each category and their respective subcategories. We also report on the total number of papers within each subcategory.
The four technical categories mentioned above have distinct visualization and sonification tags, which are represented in the left and right boxes in the ``Num.'' column.
The luminance of the boxes encodes the total number of papers in each subcategory.

\subsection{Thematic Corpus Overview}
\label{sec: thematic corpus overview}

\new{Before we employ the classification system described in \autoref{tab:tag-definitions}, we want to provide a thematic corpus overview. This overview is intended for readers who are interested in a special field of application, such as astronomy or earth sciences. While we do not intend to provide detailed descriptions of the individual articles at this point, we will present more detailed descriptions in the later subsections \ref{sec: Purpose} to \ref{sec: Users and Goals}.}

Scanning our database, we identified seven \textbf{astronomy} related papers. Riber~\cite{riber_2018_PlanethesizerApproachingExoplanet} presented a prototypical virtual and interactive audio synthesizer called \href{https://archive.org/details/PlanethesizerWindows}{\textit{Planethesizer}} that enables its users to design sonifications, especially focused on planetary data. \href{https://archive.org/details/SonifigrapherMacOSX}{\textit{Sonifigrapher}}~\cite{garciariber_2019_SonifigrapherSonifiedLight} is a virtual synthesizer that lets users sonify the light curves data from NASA's exoplanet archive. Also, the recently presented \href{https://starsounder.space/}{\textit{Sonified Hertzsprung-Russel Diagram}}~\cite{huppenkothen_2023_SonifiedHertzsprungRussellDiagram} sonifies the light curves, with the diagram acting as both the visualization and the interface to choose a star to be sonified. With this design, hearing a constant pitch will inform a user about the rotation of a star. The rotation, temperature, and other parameters of planets in our solar system were also sonified by Elmquist et al.~\cite{elmquist_2021_OpenspaceSonificationComplementing} in \href{https://vimeo.com/528822742}{\textit{OpenSpace Sonification}}. Their design can be used both with conventional computer desktop environments and in planetarium settings, and they are tailored towards public outreach and science communication. Public outreach is also the core of the publication \href{https://www.youtube.com/watch?v=5HS3tRl2Ens}{\textit{Audio Universe}} by Harrison et al.~\cite{harrison_2022_AudioUniverseTour}. The publication describes the design of a 35-minute audiovisual show about the solar system integrating visualization and sonification, as well as an audiovisual animation displaying the stars in the same order they appear to our eyes during dusk. Similarly, Russo and Santaguida~\cite{russo_2022_5000ExoplanetsListen} collaborated with NASA, celebrating the discovery of the 5000th exoplanet. \href{https://exoplanets.nasa.gov/resources/2321/5000-exoplanets-listen-to-the-sounds-of-discovery-360-video/}{Their design} displays the exoplanets as they were discovered over the years. Recently, Traver presented another audiovisual installation where users can control the auditory representation of the planets using a Midi controller~\cite{traver_2023_HarmonicesSolarisSonification}.

We identified six \textbf{medicine and health} related topics in our database, out of which three are related to brain scans~\cite{gomez_2011_DATASONIFICATIONAPPROACH,roginska_2013_ExploringSonificationAugmenting,gionfrida_2016_TripleToneSonification}, two are audiovisually displaying blow flow and aneurysm models~\cite{macdonald_2018_DataDrivenSonificationCFD,temor_2021_PerceptuallymotivatedSonificationSpatiotemporallydynamic}, and one is concerned with Covid-19 data~\cite{lemmon_2023_MappingEmergencyDesigning}.

Our database holds four idioms that we relate to \textbf{molecular science}. In an idiom presented by Rau et al.~\cite{rau_2015_EnhancingVisualizationMolecular}, scientific visualization of a molecular simulation is enhanced using parameter mapping sonification and auditory icons. Among other things, their design guides the attention of a user towards visually occluded phenomena using sonification. Ballweg et al.~\cite{ballweg_2016_InteractiveSonificationStructural} use sonification with the intention of supporting chemists and structural biologists with drug design. For their sonification plug-in, they focused on tasks that were not well supported visually in a software for the interactive visualization of molecular structures called ``UCSF Chimera.''
In the context of biomolecules simulation, Arbon et al.~\cite{arbon_2018_SonifyingStochasticWalks} developed a \href{https://vimeo.com/255391814}{sonification} displaying characteristics of the ``free energy landscape,'' a map used to study the properties of biomolecular systems. Their technique allows a user to visually inspect the physical configuration of a biomolecule while listening to their corresponding free energy landscape. Exploring the possibilities of 3D sound, Bouchara and Mones~\cite{bouchara_2020_ImmersiveSonificationProtein} suggested a work-in-progress immersive sonification model to study protein surfaces. 

The \textbf{earth science} cluster in our survey data holds eleven publications covering topics that range from oceanographic data~\cite{svoronos-kanavas_2022_ExploratoryUseAudiovisual}, wildfires~\cite{han_2022_FutureRedVisualizing}, hurricanes~\cite{ballora_2015_TwoExamplesSonification}, and climate change~\cite{bearman_2011_UsingSoundRepresent}, to sonophenology~\cite{ness_2012_Sonophenology}, seismology~\cite{holtzman_2014_SeismicSoundLab,winters_2015_SonificationTohokuEarthquake,matsubara_2016_CollaborativeStudyInteractive,pate_2022_CombiningAudioVisual}, and geospatial data displays~\cite{bearman_2012_UsingSoundRepresent,gune_2018_GraphicallyHearingEnhancing}. 

The category of \textbf{domain agnostic data display} idioms in our survey data holds 15 papers. These idioms are not designed to support users from a specific domain but are implementations tackling problems across multiple domains. Six of the papers describe software frameworks that are intended to help people design sonifications along with visual representations of their data~\cite{phillips_2019_SonificationWorkstation,peng_2023_SirenCreativeExtensible,lindetorp_2021_SonificationEveryoneEverywhere,delavega_2022_SonoUnoWebInnovative,cantrell_2021_HighchartsSonificationStudio,kondak_2017_WebSonificationSandbox}. Their unifying core goal is the democratization of sonification as a technique to represent data, hence making it accessible to more people, both professionals as well as domain experts. Other publications focus on basic research combining different sonification techniques such as parameter mapping sonification or model-based sonification with basic information visualizations such as scatterplots~\cite{enge_2022_MultimodalExploratoryData,ronnberg_2016_InteractiveSonificationVisual,yang_2017_ModeExplorerUsing}, parallel coordinates plots~\cite{bru_2023_LineHarpImportanceDriven,ronnberg_2016_InteractiveSonificationVisual}, or line charts~\cite{ferguson_2012_NavigationInteractiveSonifications,fitzpatrick_2018_StreamSegregationUtilizing} to study the potentials of audiovisual display idioms.
Groppe et al. ~\cite{groppe_2021_SoundDatabasesSonification} studied network visualization and sonification through \href{https://www.ifis.uni-luebeck.de/~groppe/soundofdatabases}{their design}, while Malikova et al.~\cite{malikova_2019_VisualauditoryVolumeRendering} show the potential of sonification to help users identify smallest symmetry differences in scalar fields visualizations. While most studies focus on metrics such as precision, error rates, or task completion times, the study by Du et al.~\cite{du_2018_ExploringRoleSound} explicitly investigates the sonification's influence on user engagement. 

\revized{Finally, we want to provide a brief overview of the remaining \textbf{other} 14 publications not part of the above thematic clusters. The topics related to the publications tagged as ``other'' are diverse, ranging from a multimodal system for analyzing business process execution data~\cite{hildebrandt_2016_CombiningSonificationVisualization} or Git version control data~\cite{north_2016_UnderstandingGitHistory}, to an audiovisual representation of the Portuguese consumption patterns~\cite{macas_2018_ConsumptionRhythmMultimodal}, to a multimodal implementation of ``Game of Life''
\cite{kariyado_2021_AuralizationThreeDimensionalCellular}, and a musical sonification aimed at conveying information about running data and emotion~\cite{ronnberg_2021_SonificationConveyingData}.
Several publications focus on audiovisual data representation in virtual or extended realities~\cite{chabot_2017_ImmersiveVirtualEnvironment,papachristodoulou_2014_SonificationLargeDatasets,papachristodoulou_2015_AugmentingNavigationComplex,berger_2019_CombiningVRVisualization,joliat_2013_SpatializedAnonymousAudio}
and two publications focus on explainable AI~\cite{lyu_2021_AIiveInteractiveVisualization,herrmann_2020_VisualizingSonifyingHow}. 
Alonso et al.~\cite{alonso-arevalo_2012_CurveShapeCurvature} presented an interface for product design that communicates a virtual object's geometrical shape using visual, haptic, and auditory stimuli, and Adhitya and Kuuskankare~\cite{adhitya_2011_SonifiedUrbanMasterplan} proposed a prototype that offers a sonification-based approach to urban design planning.}

As we have \revized{now presented a thematic overview} of all 57 papers that are part of our database, we can focus on the meta-level classification in the following subsections. With these more high-level descriptors, we intend to provide a number of versatile perspectives on the literature. They will help us identify research gaps and opportunities for the systematic integration of sonification and visualization for the future work of both research communities in \autoref{sec: discussion}. Along the discussion of these perspectives, \revized{we will present selected papers that are representative examples for the respective category.} By doing so, we intend to provide the reader with (1) a broad exploration of the field overall and (2) insights into the content of the actual papers themselves.

\newcommand{\categoryicon}[1]{\raisebox{-0.2\height}{\includegraphics[height=\baselineskip]{figures/icons/icon-#1.png}}}
\newcommand{\smalltableicon}[1]{\raisebox{-0.2\height}{\includegraphics[height=3.5mm]{figures/icons/icon-#1.png}}}
\newcommand{\statbox}[2]{\noindent\raisebox{0.7\height}{\fcolorbox[rgb]{0,0,0}{#1}{\scriptsize~}}}
\newcommand{\statboing}[1]{\noindent\raisebox{0.7\height}{\fcolorbox[hsb]{0,0,0}{0.05,#1,0.99}{\scriptsize~}}}
\newcommand{\statboingblue}[2]{\noindent\raisebox{0.7\height}{\fcolorbox[hsb]{0,0,0}{0.55,#1,0.75}{\scriptsize~}}} 
\newcommand{\statboinggreen}[2]{\noindent\raisebox{0.7\height}{\fcolorbox[hsb]{0,0,0}{0.49,#1,0.76}{\scriptsize~}}} %
\newcommand{\statboxrgb}[1]{\noindent\raisebox{0.7\height}{\fcolorbox[rgb]{0,0,0}{#1}{\scriptsize~}}}

\newcommand{\catbox}[2]{\multirow{#1}{*}{\parbox{18mm}{\raggedright \textbf{#2}}}}

\begin{table*}[]
\centering \small
\begin{tabularx}{\textwidth}{p{18mm}p{14mm}>{\raggedright}p{32mm}X}
\toprule
\textbf{Category}                       & \textbf{Num.}            & \textbf{Subcategory}             & \textbf{Explanation}                                                                                                                                                      \\ \midrule
\catbox{3}{Purpose}                     & \statboxrgb{0.4392156862745098,0.7607843137254902,0.4627450980392157}~~29      & Exploration                        & designs used for data exploration                                                                                                                                         \\
                                        & \statboxrgb{0.7215686274509804,0.8901960784313725,0.6980392156862745}~~17      & Presentation                       & designs used for data presentation                                                                                                                                        \\
                                        & \statboxrgb{0.8313725490196079,0.9333333333333333,0.807843137254902}~~11      & Both                               & designs used for both purposes above                                                                                                                                      \\
\midrule
\catbox{5}{Visualization Design}        &                        & Idiom                              & the visualization idiom, such as scatter plot or line chart                                                                                                               \\
                                        &                        & Identity channels                  & the name of the visual channel, such as color hue, or shape, that is used to communicate the identity of an item (i.e., "What" something is)          \\
                                        &                        & Magnitude channels                  & the name of the visual channel such as position or length that is used to communicate the magnitude of an attribute (i.e., "How much" something is) \\
\midrule
\catbox{5}{Sonification Design}         &                        & Technique                          & the sonification technique such as parameter mapping sonification or earcons                                                                                                           \\
                                        &                        & Identity channels                  & the auditory channel such as timbre that is used to communicate the identity of an item (i.e., "What" something is)                                                       \\
                                        &                        & Magnitude channel                  & the auditory channel such as pitch or duration that is used to communicate the magnitude of an attribute (i.e., "How much" something is)                                  \\

\midrule
\catbox{3}{Reading Level}               & \statboxrgb{0.6549019607843137,0.21176470588235294,0.011764705882352941}~50~~~\statboxrgb{0.4117647058823529,0.6745098039215687,0.8313725490196079}~~29       & Whole                              & designs that display \textit{all} datapoints                                                                                                                                       \\
                                        & \statboxrgb{0.9921568627450981,0.8274509803921568,0.6627450980392157}~13~~~\statboxrgb{0.4117647058823529,0.6745098039215687,0.8313725490196079}~~29       & Group                              & designs that display a \textit{group} of datapoints                                                                                                                                \\
                                        & \statboxrgb{0.996078431372549,0.9254901960784314,0.8549019607843137}~4\enspace~~~\statboxrgb{0.7137254901960784,0.8313725490196079,0.9137254901960784}~~17       & Single                             & designs that display \textit{single} datapoints                                                                                                                                    \\
\midrule
\catbox{4}{Search Level}                & \statboxrgb{1,0.9450980392156862,0.8901960784313725}~2\enspace~~~\statboxrgb{0.9411764705882353,0.9686274509803922,0.9921568627450981}~~2\enspace       & Lookup                             & the user knows the location and the target of the search (see \autoref{fig: search_munzner})                                                                                                                 \\
                                        & \statboxrgb{0.996078431372549,0.8784313725490196,0.7568627450980392}~9\enspace~~~\statboxrgb{0.8862745098039215,0.9333333333333333,0.9725490196078431}~~6\enspace       & Browse                             & the user knows the location but not the target of the search                                                                                                              \\
                                        & \statboxrgb{0.996078431372549,0.8980392156862745,0.8}~7\enspace~~~\statboxrgb{0.8745098039215686,0.9215686274509803,0.9686274509803922}~~7\enspace       & Locate                             & the user doesn't know the location but the target of the search                                                                                                           \\
                                        & \statboxrgb{0.8823529411764706,0.3333333333333333,0.03529411764705882}~40~~~\statboxrgb{0.1450980392156863,0.4588235294117647,0.7137254901960784}~~42       & Explore                            & the user doesn't know the location or the target of the search                                                                                                            \\
                                        & \statboxrgb{0.9921568627450981,0.8549019607843137,0.7137254901960784}~11~~~\statboxrgb{0.8588235294117647,0.9137254901960784,0.9647058823529412}~~8\enspace       & None                            &  none of the above                                                                                 \\
\midrule
\catbox{4}{Dataset Type}                & \statboxrgb{0.984313725490196,0.5411764705882353,0.22745098039215686}~29~~~\statboxrgb{0.1803921568627451,0.49411764705882355,0.7333333333333333}~~40       & Table                              & data constructed from items and attributes (spreadsheets; see \autoref{fig: datasettypes_munzner})                                                                                                                 \\
                                        & \statboxrgb{0.996078431372549,0.9176470588235294,0.8392156862745098}~5\enspace~~~\statboxrgb{0.9254901960784314,0.9568627450980393,0.9882352941176471}~~3\enspace       & Network                            & data constructed from items (nodes), links and potentially attributes                                                                                                     \\
                                        & \statboxrgb{0.996078431372549,0.8784313725490196,0.7568627450980392}~9\enspace~~~\statboxrgb{0.8745098039215686,0.9215686274509803,0.9686274509803922}~~7\enspace       & Field                              & data constructed from grids (positions) and attributes                                                                                                                    \\
                                        & \statboxrgb{0.9921568627450981,0.7098039215686275,0.4627450980392157}~20~~~\statboxrgb{0.8470588235294118,0.9058823529411765,0.9607843137254902}~~9\enspace       & Geometry                           & data constructed from items and positions                                                                                                                                 \\
\midrule
\catbox{4}{Level of Measurement}        & \statboxrgb{0.9921568627450981,0.6705882352941176,0.40784313725490196}~22~~~\statboxrgb{0.788235294117647,0.8705882352941177,0.9411764705882353}~~13       & Nominal                            & data that builds categories (such as different fruits)                                                                                                                    \\
                                        & \statboxrgb{0.996078431372549,0.8901960784313725,0.7803921568627451}~8\enspace~~~\statboxrgb{0.8588235294117647,0.9137254901960784,0.9647058823529412}~~8\enspace       & Ordinal                            & data that builds ordered categories (such as t-shirt sizes)                                                                                                               \\
                                        & \statboxrgb{0.9882352941176471,0.5803921568627451,0.27450980392156865}~27~~~\statboxrgb{0.36470588235294116,0.6431372549019608,0.8156862745098039}~~31       & Interval                           & data that has equal intervals, such as the time on the clock                                                                                                              \\
                                        & \statboxrgb{0.9882352941176471,0.5803921568627451,0.27450980392156865}~27~~~\statboxrgb{0.2980392156862745,0.596078431372549,0.792156862745098}~~34       & Ratio                              & data that has equal intervals and a meaningful zero point, such as length or weight                                                                                       \\
\midrule
\catbox{4}{Level of Redundancy}         & \statboxrgb{0.7215686274509804,0.8901960784313725,0.6980392156862745}~~17       & Redundant                          & a design mapping all displayed information to both senses redundantly (see \autoref{fig: level of redundancy})                                                                                                   \\
                                        & \statboxrgb{0.7803921568627451,0.9137254901960784,0.7529411764705882}~~14       & Complementary                      & a design mapping part of the information exclusively to the visualization and another part of the information exclusively to the sonification                           \\
                                        & \statboxrgb{0.4627450980392157,0.7725490196078432,0.47843137254901963}~~28       & Mixed                              & a design mapping some information redundantly, some information complementary                                                                                  \\
\midrule
\catbox{5}{Evaluation System}           & \statboxrgb{0.8156862745098039,0.9294117647058824,0.792156862745098}~~12       & User Performance                   & evaluations collecting metrics such as error rates or task completion times                                                                                               \\
                                        & \statboxrgb{0.8313725490196079,0.9333333333333333,0.807843137254902}~~11       & User Experience                    & evaluations collecting user feedback, typically done in a usability test                                                                                                  \\
                                        & \statboxrgb{0.9490196078431372,0.9803921568627451,0.9372549019607843}~~2       & Algorithmic Performance            & evaluations doing measurements without users such as rendering speed                                                                                                      \\
                                        & \statboxrgb{0.7215686274509804,0.8901960784313725,0.6980392156862745}~~17       & Qualitative Result Inspection      & evaluations providing subjective arguments on the quality of the result                                                                                                   \\
                                        & \statboxrgb{0.6588235294117647,0.8627450980392157,0.6352941176470588}~~20       & None                               & no evaluation                                                                                                                              \\
\midrule
\catbox{4}{Target Platform}             & \statboxrgb{0.12941176470588237,0.5411764705882353,0.26666666666666666}~~43       & Desktop Computer Display           & conventional screen on desktop computer                                                                                                                                   \\
                                        & \statboxrgb{0.9294117647058824,0.9725490196078431,0.9137254901960784}~~4       & XR                                 & extended reality settings such as virtual or augmented reality glasses                                                                                                     \\
                                        & \statboxrgb{0.8470588235294118,0.9411764705882353,0.8274509803921568}~~10       & Physical Environments & environments that foster collaboration such as a CAVE                                                                            \\
                                        & \statboxrgb{0.9176470588235294,0.9686274509803922,0.9019607843137255}~~5       & Touch Display                      & interactive screen that users can interact with via touch                                                                                  \\

\midrule
\catbox{2}{User Interaction}            & \statboxrgb{0.23921568627450981,0.6431372549019608,0.34509803921568627}~~37       & Yes                                & designs that require user interaction other than pressing ``play''                                                                                                            \\
                                        & \statboxrgb{0.6588235294117647,0.8627450980392157,0.6352941176470588}~~20       & No                                 & designs that require no user interaction                                                                                                                                  \\

\midrule
\catbox{3}{Users}                       & \statboxrgb{0.5411764705882353,0.807843137254902,0.5333333333333333}~~25       & Domain Experts                     & domain experts that are not visualization or sonification researchers                                                                                                     \\
                                        & \statboxrgb{0.8901960784313725,0.9568627450980393,0.8745098039215686}~~7       & Researchers                        & visualization or sonification researchers                                                                                                                                 \\
                                        & \statboxrgb{0.3843137254901961,0.7333333333333333,0.43137254901960786}~~31       & General Public                     & the interested general public                                                                                                                                             \\
\midrule
\catbox{4}{Goals}                       & \statboxrgb{0.8784313725490196,0.9529411764705882,0.8588235294117647}~~8       & Education                          & idioms designed for education                                                                                                                                 \\
                                        & \statboxrgb{0.2823529411764706,0.6745098039215687,0.37254901960784315}~~35       & Data Analysis                      & idioms designed for data analysis (incl. data exploration)                                                                                                    \\
                                        & \statboxrgb{0.9372549019607843,0.9764705882352941,0.9254901960784314}~~3       & Research                           & idioms designed for visualization and sonification research                                                                                                    \\
                                        & \statboxrgb{0.7215686274509804,0.8901960784313725,0.6980392156862745}~~17       & Public Engagement                  & idioms designed for public engagement                                                                                                                         \\
\midrule
\catbox{2}{Demo}                        & \statboxrgb{0.4627450980392157,0.7725490196078432,0.47843137254901963}~~29       & Yes                                & the paper links to a demo such as an interactive website, a video, or an audio recording                                                                                 \\
                                        & \statboxrgb{0.5882352941176471,0.8313725490196079,0.5764705882352941}~~23       & No                                 & the paper doesn't link to a demo                                                                                                                                         \\
                                        & \statboxrgb{0.9176470588235294,0.9686274509803922,0.9019607843137255}~~5       & Yes, but not online                & the paper provides a link that is not online anymore                                                                                                                     \\ \bottomrule
\end{tabularx}%
\caption{Descriptions and distributions of all used classification tags. For the categories with two separate tags under the ''Num." column, the left and right boxes represent the visualization and sonification distributions, respectively.}
\label{tab:tag-definitions}
\end{table*}

\subsection{Purpose}
\label{sec: Purpose}

In this section, we will discuss two different purposes an audiovisual idiom can be designed for, inspired by the taxonomy described in~\cite{munzner_visualization_2015}: exploration and presentation. The \textit{purpose} should be read as the general design goal of a tool with a broad perspective. Therefore, the term exploration also covers what is widely referred to as data analysis. 

\subsubsection{Exploration}
A wide range of audiovisual idioms in our database support users with the exploration of data (see \autoref{tab:tag-definitions}). We refer to the purpose of exploration in cases where a user intends to acquire new knowledge from their data by using an audiovisual display idiom. 

Exploratory data analysis is the core aim of model-based sonification~\cite{hermann2002sonification}. The \href{https://pub.uni-bielefeld.de/record/2920473}{\textit{Mode Explorer}} is an interactive audiovisual display idiom, combining a scatterplot visualization and model-based sonification to explore high-dimensional data~\cite{yang_2018_InteractiveModeExplorer}. To explore the data, a user can ``scratch'' a scatter plot of dimensionality-reduced data with an interactive pen. The scratching introduces a virtual particle to the high-dimensional space, which will follow a gradient descent to the nearest mode in the data. A user will be able to hear the ``kinetic energy'' of this virtual particle. While the particle travels closer and closer to its final mode, the sound gradually turns more and more harmonic, finally resulting in a clear pitch when a mode is reached. 

\href{https://www.zhuoyuelyu.com/aiive}{\textit{AIive}} combines visualization and sonification in a virtual reality environment~\cite{lyu_2021_AIiveInteractiveVisualization}. The idiom helps users understand the basic concepts of neural networks. Users can manipulate the weights of the nodes of a virtually displayed 3D neural network by dragging the nodes in 3D space. In real-time, the sonification displays the loss and accuracy of the neural network, therefore enabling a user to explore different constellations of node weights. Users are also able to add or delete nodes, hence experimenting with the complexity of the neural network. While the authors do not provide a user study in their paper, it is most plausible the exploratory character of the design can support people in understanding the basics of neural networks. 

In a recent publication, Lemmon et al. presented an \href{https://ericlemmon.net/ison2022-demo-video/}{audiovisual map} idiom that seeks to tackle some of the sociotechnical challenges associated with epidemiological mapping~\cite{lemmon_2023_MappingEmergencyDesigning}. Using their idiom, users can interactively explore Suffolk County's experience with Covid-19. The black population and associated case numbers are displayed on the left audio channel, while the white population and associated case numbers are displayed on the right channel via the pitch of sine and triangle waves. A correlation between black and white population numbers and COVID-19 case numbers and their dependency on different regions becomes clearly apparent when listening to the sonification while brushing the map using the computer mouse.

\subsubsection{Presentation}
We refer to the purpose of presentation when a user intends to present prior knowledge to others or in cases where a user is presented with information that is new to them but not to the designer.

One potential application of a presentation-only approach is in citizen science, exemplified by the communication of significant discoveries, such as NASA's announcement of the 5000th exoplanet~\cite{russo_2022_5000ExoplanetsListen}. This representation is explicitly crafted for communication with the general public, primarily through social media channels. Translating data into various sensory modalities is straightforward, while significant attention is devoted to achieving a pleasing and harmonious aesthetic. The sonification, complemented by two animated videos, enhances visual comprehension, vividly portraying celestial spheres with dynamic elements representing exoplanet characteristics. NASA's strategic dissemination on social media platforms garnered substantial engagement, with positive feedback indicating the presentation's broad appeal. 

The presentation objective appears linked to eliciting emotional engagement from end users, as evident in Rönnberg's research~\cite{ronnberg_2021_SonificationConveyingData}. \href{https://vimeo.com/397235072}{This representation} involves assessing running statistics, weather data, and associated emotions. Presented as an animated visualization synchronized with sonification, it depicts weekly runs emphasizing evoking emotional responses rather than precise data interpretation—a practice denoted by the author as ``musification'' (see further discussion about sonification and music in Vickers, 2016~\cite{vickers2016sonification}), leveraging sound's dual capacity to illustrate data and engender a musical experience. The study assesses the representation's efficacy through a user study, wherein participants watched a video. Results indicate that the sonification effectively conveys intended emotions but at the expense of a less accurate data representation.

Recent works delve into environmental concerns and the promotion of awareness. One such noteworthy contribution is presented in~\cite{han_2022_FutureRedVisualizing}. This publication centers around a multimodal museum installation designed to foster public engagement with wildfire forecasts in specifically chosen California and South Korea regions. The project incorporates interactive data visualization, sonification, and 3D-printed sculptures. Combining these elements conveys wildfire data, creating a comprehensive and immersive experience. The installation allows audiences to explore the representation freely through contactless interaction. Without user engagement, the wildfire data representation seamlessly loops across the screen. Showcased at the ARKO Art Center in Seoul, this project successfully captivated the attention of 20 users. The outcomes revealed a heightened level of interest and comprehension regarding the impact of wildfires through the effective utilization of multimodal interaction. The project emphasizes observations and prompts audience reflection, leveraging new media to enhance public climate change comprehension.

\subsubsection{Exploration and Presentation}

Some of the designs we identified were designed both to explore and to present data. Inclusion in this category requires the designer to put similar weight on the exploration and presentation of the data. 

To present a few examples, Paté et al.~\cite{pate_2022_CombiningAudioVisual} demonstrate their \href{https://parthurp.github.io/homepage/SpatialSeismicSoundscapes_article2021.html}{audiovisual display idiom} with three seismic data sets, where the sonification methods are adapted to the specific properties of each data set. A multi-scale audification method is presented where different speed factors are used depending on the size of each dataset, and different sound designs are used to highlight specific properties of the data.
Huppenkothen et al.~\cite{huppenkothen_2023_SonifiedHertzsprungRussellDiagram} created an audiovisual version of the \href{https://starsounder.space/}{Hertzsprung-Russell Diagram} where the user can listen and compare auditory representations of each type of star that is included in the diagram (see \autoref{fig: platforms}(c)). It is also possible to filter the diagram according to certain criteria, which, in turn, filters which sonifications can be listened to.
Elmquist et al.~\cite{elmquist_2021_OpenspaceSonificationComplementing} created complementing sonifications of the planets in the solar system, which were integrated with a visualization, where a user can explore the properties of the planets and make comparisons between them. The user can listen to all of the sonifications for each planet at the same time or enable specific sonifications for each planet to compare specific properties. 
Du et al.~\cite{du_2018_ExploringRoleSound} conducted a study investigating the enhanced visualization of basketball player movement data during a game. The visualization is designed to convey the offensive and defensive dynamics of a team. The interface primarily enables users to specify a particular time range of interest, providing a more comprehensive view of information within that timeframe. Notably, sonification is integrated exclusively during the exploration phase. 

\subsection{Audiovisual Idiom Design}
\label{sec: Audiovisual Idiom design}

Designers of audiovisual idioms have \revized{countless} possibilities to combine their visualizations and sonifications. They can choose from a vast number of established visualization idioms and sonification techniques. Having chosen two designs to integrate with each other, they are free to choose the visual and auditory channels they want to map their data to. During the classification of the surveyed literature, we used six different categories to capture the state-of-the-art of audiovisual idiom design. The categories are the visualization idiom, the sonification technique, the visual identity- and magnitude channels, as well as the auditory identity- and magnitude channels. The constructs of identity- and magnitude channels are used to distinguish between encodings that communicate ``what'' something is and ``how much of'' something there is~\cite{munzner_visualization_2015,enge_2023_unified}. Typical visual identity channels are the shape of a visual mark or its color hue. Typical visual magnitude channels are the position or the length of a visual mark. A typical auditory identity channel is the timbre of a sound, often generated using different musical instruments. Typical auditory magnitude channels are pitch and loudness~\cite{dubusSystematicReviewMapping2013, caiola_2022_AudiovisualSonificationsDesign}. In the following, we will make use of these descriptions to shed light on some of the designs in our database.

\begin{figure}
    \centering
    \includegraphics[width=0.9 \linewidth]{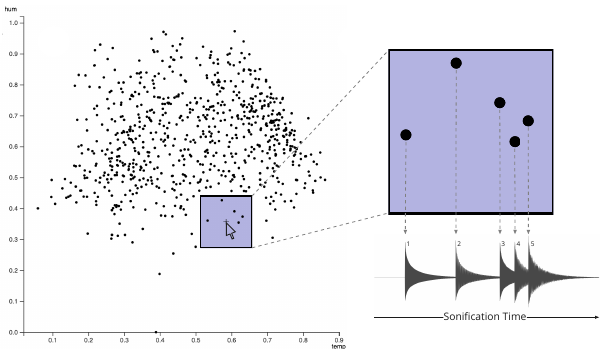}
    \caption{The SoniScope audiovisual display idiom~\cite{enge_2022_MultimodalExploratoryData}. Users can interactively select a region in a scatterplot to sonify an additional data attribute for the respective items.}
    \label{fig: soniscope}
\end{figure}

\href{https://phaidra.fhstp.ac.at/detail/o:4831}{\textit{SoniScope}}~\cite{enge_2022_MultimodalExploratoryData} is an interactive audiovisual display idiom that combines a scatterplot visualization with an interactive parameter mapping sonification (compare \autoref{fig: soniscope}). Inspired by the technique of auscultation, the SoniScope acts as a stethoscope for data, providing an auditory lens to ``listen into'' one's data. The exploratory character of the idiom is most apparent when the sonification displays non-visual data dimensions. To interact with the idiom, a user can brush the data using a visual lens. Clicking into the scatterplot will then trigger the sonification of a non-visual attribute of the selected data items. Regarding the used magnitude channels, we see a common combination of visual position and the pitch of the sounds, also employed in publications such as~\cite{bearman_2012_UsingSoundRepresent,delavega_2022_SonoUnoWebInnovative,ferguson_2012_NavigationInteractiveSonifications,fitzpatrick_2018_StreamSegregationUtilizing,kondak_2017_WebSonificationSandbox}. 

Bearman~\cite{bearman_2011_UsingSoundRepresent} studied the possibilities of displaying the uncertainty in future climate projections for the UK. The proposed audiovisual idiom shows a map of the UK with a heat map overlay displaying the projected temperature values. This visualization is combined with the sonification technique of interactive parameter mapping. People in a user study were asked to hover over the heat map with a mouse, triggering the sonification of the respective region. A higher pitch of a trumpet sound communicated a higher uncertainty for that region. Hence, for their design, Bearman used the spatial position as a visual identity channel and color as a visual magnitude channel. \href{https://vimeo.com/17029358}{This sonification} did not use an identity channel (only one sound could be heard) but used pitch as a magnitude channel.

\begin{figure*}
    \centering
    \includegraphics[width=\linewidth]{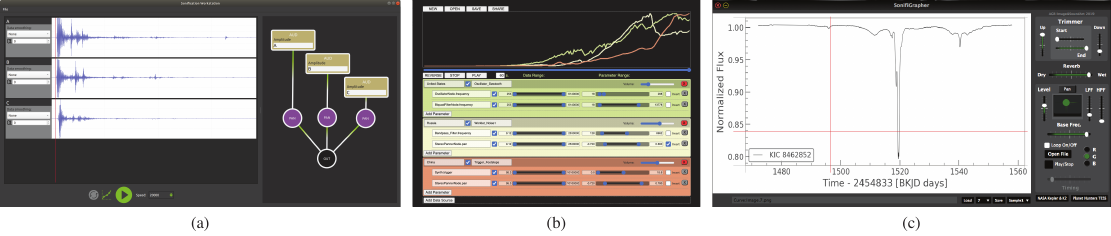}
    \caption{Examples of frameworks for sonification in combination with visualization. (a) The Sonification Workstation by Phillips and Cabrera~\cite{phillips_2019_SonificationWorkstation} is designed for general sonification tasks. (b) The \href{https://hanslindetorp.github.io/SonificationToolkit/}{\textit{WebAudioXML Sonification Toolkit}} (WAST) user interface by Lindetorp and Falkenberg~\cite{lindetorp_2021_SonificationEveryoneEverywhere} also aimed for general sonification tasks. (c) The Sonifigrapher virtual synthesizer interface by Riber~\cite{garciariber_2019_SonifigrapherSonifiedLight} for sonifying light curves. All screenshots CC BY 4.0.}
    \label{fig: frameworks}
\end{figure*}

Several papers in our database describe frameworks that are explicitly developed for the design of sonifications (see examples in \autoref{fig: frameworks}), always in combination with a visualization~\cite{phillips_2019_SonificationWorkstation,peng_2023_SirenCreativeExtensible,lindetorp_2021_SonificationEveryoneEverywhere,delavega_2022_SonoUnoWebInnovative, garciariber_2019_SonifigrapherSonifiedLight,cantrell_2021_HighchartsSonificationStudio}. As a representative and flexible example we want to discuss the \href{https://sonification.highcharts.com/#/}{\textit{Highcharts Sonification Studio}}~\cite{cantrell_2021_HighchartsSonificationStudio}, a collaboration between the company \href{https://www.highcharts.com/}{Highcharts} and the \href{http://sonify.psych.gatech.edu/}{Georgia Tech Sonification Lab}. Regarding the design of the visualization, the environment offers line- and area charts, as well as scatter plots, bar charts, and pie charts, all with their respective standard visual channels such as position, length, angle, or color hue. The sonification is done using parameter mapping with several options for auditory magnitude channels such as pitch, spatial position in the stereo field, loudness, or harmonic range. Different attributes in the data can be distinguished using different musical instruments, hence the employed auditory identity channel is the timbre of the different sounds. The default option of the Highcharts sonification studio is set to combine a line chart with an auditory graph, which traditionally plays back the line from left to right using pitch over time. 

In their paper, Winters et al.~\cite{winters_2015_SonificationTohokuEarthquake} describe the design of the visualization and the sonification of the 2011 Tohoku earthquake in Japan. 
The design combines four line charts of four seismographic recording stations in Japan with an audification of the same data streams. The essence of the audification is that the low-frequency recordings of the earthquake are played back at a faster tempo, making them audible to the human ear. This very direct mapping between the physical (non-audible) phenomenon and the audification to the audible range results in a rich auditory impression that becomes informative in an ecologically valid manner. The paper mostly discusses the popularity of this sonification on \href{https://www.youtube.com/watch?v=3PJxUPvz9Oo}{YouTube}, currently with around 90k views, explicitly mentioning the relevance of combining the sonification with visualization for its success in public outreach. The visualization uses position as both its identity and its magnitude channel. The sonification employs the spatial stereo position of the four channels to distinguish between them, hence as their identity channel. The magnitude channels are a mix of pitch and timbre, resulting from the direct mapping between the physical phenomenon and the sound.

\noindent
\textbf{Discussion:}
From the data, we see that parameter mapping is by far the most used sonification technique for the design of audiovisual idioms -- 53 out of 57 papers in the database employ some sort of parameter mapping sonification. Only seven of them combine parameter mapping sonification with other techniques such as audification~\cite{matsubara_2016_CollaborativeStudyInteractive,huppenkothen_2023_SonifiedHertzsprungRussellDiagram}, earcons~\cite{hildebrandt_2016_CombiningSonificationVisualization, north_2016_UnderstandingGitHistory}, or auditory icons~\cite{ballweg_2016_InteractiveSonificationStructural,rau_2015_EnhancingVisualizationMolecular,papachristodoulou_2015_AugmentingNavigationComplex}. The technique of audification is part of the database three times on its own~\cite{pate_2022_CombiningAudioVisual,holtzman_2014_SeismicSoundLab,winters_2015_SonificationTohokuEarthquake}, and once in combination with earcons in a second design of a paper~\cite{pate_2022_CombiningAudioVisual}. Yang and Hermann are the only authors employing model-based sonification with their \textit{Mode Explorer} design~\cite{yang_2017_ModeExplorerUsing}. 
The most prominent auditory identity channel in our literature corpus is timbre (used 21 times), and the most used auditory magnitude channel is pitch (47 times), also used in combination with other channels. These findings are just in line with other meta-analyses of the field of sonification~\cite{caiola_2022_AudiovisualSonificationsDesign,dubusSystematicReviewMapping2013}.
The visualizations in our corpus employ idioms such as line charts (10 times), scatter plots (8 times), and maps (7 times), as well as networks, bar charts, heatmaps, and several other idioms. These publications use the identity channels color hue (27 times), position (11 times), as well as shape (11 times). The employed visual magnitude channels are position (33 times), color hue (15 times), size (9 times), and other channels such as opacity, tilt, or animation.

Tagging the surveyed literature made us realize that using the concept of the ``channel'' for techniques such as audification or model-based sonification is not trivial. With these techniques, the designer of a sonification, to some extent, loses control over its sonic outcome. This contradicts the definition of a visual channel as ``a way to control the appearance of marks''~\cite{munzner_visualization_2015}. Instead, thinking of the ``channel'' as ``the quality of a mark that transports information,'' also allows for the description of audifications and model-based sonifications. While these techniques typically result in highly complex sound sequences, it will often be qualities such as pitch or timbre that are informative to the listener. In general, we see a quite diverse field of different audiovisual idiom designs in our corpus.

\subsection{Reading Levels}
\label{sec: Reading Levels}

\begin{table}[]
\centering
\resizebox{0.6\columnwidth}{!}{%
\begin{tabular}{@{}c|ccc@{}}
\toprule
\multirow{2}{*}{\textbf{Sonification}} & \multicolumn{3}{c}{\textbf{Visualization}} \\
                                    & whole   & group   & single  \\ \cmidrule(l){2-4} 
whole                               & 27      & 3         & 1              \\
group                                & 24       & 12         & 1              \\
single                                & 14       & 4          & 3              \\ \bottomrule
\end{tabular}%
}
\caption{The number of entries for Bertin's different reading levels~\cite{bertin_1983_semiology} and their distribution to the two display types.}
\label{tab: Readingleveltable}
\end{table}

The classification of designs with respect to reading levels is inspired by the taxonomy by Bertin in his seminal book Semiology of Graphics~\cite{bertin_1983_semiology}. The three reading levels describe the amount of data a user studies using a specific tool. The ``Whole'' level describes tools that enable the user to ask questions about ``all'' of the data under consideration. The ``Group'' level describes tools that enable the user to ask questions about a ``subgroup'' of the data under consideration, and the ``Single'' level describes tools that enable the user to study ``single'' items. For this category, we decided to assign two tags per entry in our database: One for the visualization part and one for the sonification part of the idiom. The classification using reading levels aims towards potential differences in the distribution of tasks between sonification and visualization, such as overview and detail phenomena. 

Rau et al.~\cite{rau_2015_EnhancingVisualizationMolecular} presented an audiovisual idiom that lets users interactively explore molecular structures by using a ``virtual microphone'' that can be placed inside a 3D molecular visualization. The visual design provides an overview using the ``whole'' level while displaying details at the ``group'' level via the sonification. At the same time, the design provides a user with information about potentially visually occluded data. Hence, it makes the user aware of the existence of a phenomenon that they could, if relevant, study in detail at a later point. 
The metaphor of the microphone to be used to ``listen into the data'' is similar to the one of the stethoscope in~\cite{enge_2022_MultimodalExploratoryData}, where the visualization provides an overview while users can interactively choose a subset of data to display acoustically.
An example of an idiom working at the sonification ``whole'' level and the visualization ``group'' level is the \textit{Mode Explorer}~\cite{yang_2017_ModeExplorerUsing}. The design sonifies a high-dimensional data space while visualizing a two-dimensional projection of that space with a scatterplot. 

\noindent
\textbf{Discussion:}
Out of all possible combinations of reading levels and display type, the majority of cases (27) use the ``Whole'' level for both the sonification and the visualization (some of them are~\cite{elmquist_2021_OpenspaceSonificationComplementing, macas_2018_ConsumptionRhythmMultimodal, russo_2022_5000ExoplanetsListen, svoronos-kanavas_2022_ExploratoryUseAudiovisual, winters_2015_SonificationTohokuEarthquake}). The next biggest group in our database holds 24 papers that can display data at the ``Whole'' level using their visualization and at the ``group'' level using their sonification, such as~\cite{bearman_2012_UsingSoundRepresent, bru_2023_LineHarpImportanceDriven, ferguson_2012_NavigationInteractiveSonifications, gune_2018_GraphicallyHearingEnhancing, rau_2015_EnhancingVisualizationMolecular}. Generally speaking, we can observe that the visualization parts are mostly covering the same or a higher level of reading than the sonification parts of a design (compare \autoref{tab: Readingleveltable}). This phenomenon can be directly related to the distribution of tasks on the two display types, with visualization more often providing an overview and the sonification rather displaying details. In \autoref{sec: discussion}, we will argue for breaking such patterns in the future work of the community. 

\subsection{Search Level}
\label{sec: Search Levels}

\begin{figure}
    \centering
    \includegraphics[width=\linewidth]{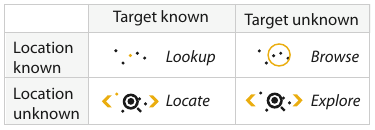}
    \caption{The four search levels suggested by Munzner: lookup, browse, locate, and explore. (Figure from “Visualization Analysis and
Design”~\cite{munzner_visualization_2015} by Tamara Munzner, with illustrations by Eamonn Maguire, CC BY 4.0.)}
    \label{fig: search_munzner}
\end{figure}

\begin{table}[]
\centering
\resizebox{0.7\columnwidth}{!}{%
\begin{tabular}{@{}c|cccc@{}}
\toprule
\multirow{2}{*}{\textbf{Sonification}} & \multicolumn{4}{c}{\textbf{Visualization}} \\
                                    & explore   & browse   & locate  & lookup  \\ \cmidrule(l){2-5} 
explore                               & 37      & 4         & 5        & 5       \\
browse                                & 3       & 2         & 2        & 2       \\
locate                                & 4       & 3          & 5        & 5        \\
lookup                                & 1       & 0          & 1        & 2       \\ \bottomrule
\end{tabular}%
}
\caption{The number of entries for Munzner’s different search levels~\cite{munzner_visualization_2015} and their distribution to the two display types.}
\label{tab: search levels}
\end{table}

Inspired by the taxonomy suggested by Munzner~\cite{munzner_visualization_2015}, we tagged four different search levels, again assigned individually for the sonification and the visualization parts of the idiom.
The type of search a user applies depends on their prior knowledge (see \autoref{fig: search_munzner}). Users who know what they are looking for and where they can find it will do a \emph{lookup}. Searching for an unknown pattern at a known location is called \emph{browsing} while \emph{locating} is a search without knowing the location but the pattern one is looking for. A search uninformed both with regards to the location and the type of pattern one is looking for is called \emph{exploration}. Designs that combine sonification and visualization could serve each of the described search levels as well as a combination of them. 

Malikova et al.~\cite{malikova_2019_VisualauditoryVolumeRendering}, for example, presented an \href{https://vimeo.com/323545930}{idiom} that helps users explore scalar fields. The regional magnitude of small areas of a scalar field is sonified using pitch. From the search level perspective, a user of this idiom does not generally know what to look or listen for at the beginning of the analysis, and therefore, both modalities are used to explore the presented data. The system could, for example, reveal the existence of small symmetry differences in scalar fields. What visually seems to be a perfectly symmetrical field could become apparent as not quite symmetric when two similar pitches result in an acoustic phenomenon called ``frequency beating.'' The clearly audible phenomenon makes a listener aware of the non-symmetry of the field, potentially resulting in them taking a closer look at their data. 

\noindent
\textbf{Discussion:}
Generally speaking, many idioms can be used for more than one search level, and which one they are used for seems to be dependent on the user's prior knowledge (both about the data and about the idiom). Nevertheless, the majority of papers in the surveyed literature offer the search level of ``explore'' and, in general, most papers use the same search level for both their visualization and their sonification parts (see \autoref{tab:tag-definitions} and \autoref{tab: search levels}). Some examples of designs employing the same search level are~\cite{chabot_2017_ImmersiveVirtualEnvironment, du_2018_ExploringRoleSound, enge_2022_MultimodalExploratoryData, ferguson_2012_NavigationInteractiveSonifications, garciariber_2019_SonifigrapherSonifiedLight, gomez_2011_DATASONIFICATIONAPPROACH, han_2022_FutureRedVisualizing}, while fewer combine different search levels for the two modalities~\cite{hildebrandt_2016_CombiningSonificationVisualization, rau_2015_EnhancingVisualizationMolecular, ronnberg_2016_InteractiveSonificationVisual, adhitya_2011_SonifiedUrbanMasterplan, ballora_2015_TwoExamplesSonification, cantrell_2021_HighchartsSonificationStudio}.

\subsection{Dataset Type and Level of Measurement}
\label{sec: dataset types and level of measurement}

\begin{figure}
    \centering
    \includegraphics[width=1\linewidth]{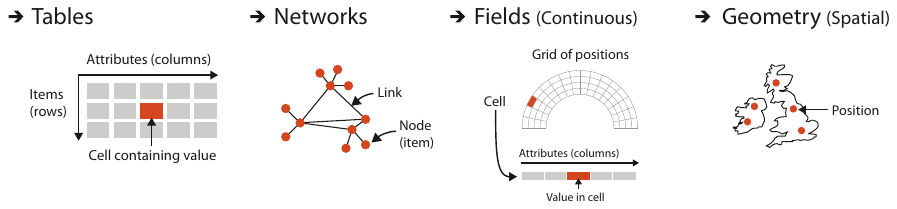}
    \caption{Four dataset types suggested by Munzner: tables, networks, fields, and geometries. (Figure from ``Visualization Analysis and Design''~\cite{munzner_visualization_2015} by Tamara Munzner, with illustrations by Eamonn Maguire, CC BY 4.0.)}
    \label{fig: datasettypes_munzner}
\end{figure}

\begin{table}[]
\centering
\resizebox{0.7\columnwidth}{!}{%
\begin{tabular}{@{}c|cccc@{}}
\toprule
\multirow{2}{*}{\textbf{Sonification}} & \multicolumn{4}{c}{\textbf{Visualization}} \\
                                       & table   & network   & field   & geometry   \\ \cmidrule(l){2-5} 
table                                  & 23      & 1         & 4       & 12         \\
network                                & 0       & 2         & 0       & 0          \\
field                                  & 0       & 0         & 7       & 1          \\
geometry                               & 1       & 0         & 0       & 9          \\ \bottomrule
\end{tabular}%
}
\caption{The number of entries for the different dataset types~\cite{munzner_visualization_2015} and their combinations between the two display techniques. Notably, the visualization of geometries is often combined with the sonification of table data.}
\label{tab: Dataset types}
\end{table}

\begin{table}[]
\centering
\resizebox{0.7\columnwidth}{!}{%
\begin{tabular}{@{}c|cccc@{}}
\toprule
\multirow{2}{*}{\textbf{Sonification}} & \multicolumn{4}{c}{\textbf{Visualization}} \\
                                       & ratio   & interval   & ordinal  & nominal  \\ \cmidrule(l){2-5} 
ratio                                  & 21      & 11         & 0        & 10       \\
interval                               & 7       & 22         & 2        & 12       \\
ordinal                                & 0       & 2          & 8        & 7        \\
nominal                                & 4       & 2          & 5        & 12       \\ \bottomrule
\end{tabular}%
}
\caption{The number of entries for the different levels of measurement and their combinations between the two display techniques. Most idioms use identical levels of measurement. Notably, 42 cases use a higher level of measurement with their sonification than with their visualization (numbers above the main diagonal). Twenty cases use the levels of measurement the other way around (numbers under the main diagonal).}
\label{tab: level of measurement}
\end{table}

Two more categories that we tagged individually for the sonification parts and the visualization parts are the dataset type (see \autoref{fig: datasettypes_munzner} and \autoref{tab:tag-definitions}) and the level of measurement (see \autoref{tab:tag-definitions}). We distinguish between the four dataset types of tables, networks, fields, and geometries~\cite{munzner_visualization_2015} and between the four levels of measurement of nominal, ordinal, interval, and ratio scale.
 
It is reasonable that most idioms display data from identical dataset types with both their visualization and their sonification. The most prominent dataset type is the table, used in designs such as~\cite{bru_2023_LineHarpImportanceDriven, delavega_2022_SonoUnoWebInnovative, du_2018_ExploringRoleSound, enge_2022_MultimodalExploratoryData, ferguson_2012_NavigationInteractiveSonifications, hildebrandt_2016_CombiningSonificationVisualization, macas_2018_ConsumptionRhythmMultimodal, ronnberg_2016_InteractiveSonificationVisual}. \autoref{tab: Dataset types} reveals a notable exception from the dominance of identical dataset types: the combination of visualized geometry in combination with a sonified tables, often related to maps and supplemented regional information~\cite{bearman_2012_UsingSoundRepresent, lemmon_2023_MappingEmergencyDesigning, matsubara_2016_CollaborativeStudyInteractive, ness_2012_Sonophenology, russo_2022_5000ExoplanetsListen}.

Concerning the levels of measurement of the displayed data, we observe, again, that most idioms display data from the same level (see \autoref{tab: Dataset types}). We list a selection of cases for ratio data ~\cite{alonso-arevalo_2012_CurveShapeCurvature, du_2018_ExploringRoleSound, gionfrida_2016_TripleToneSonification, pate_2022_CombiningAudioVisual, ronnberg_2021_SonificationConveyingData}, for interval data~\cite{berger_2019_CombiningVRVisualization, delavega_2022_SonoUnoWebInnovative, garciariber_2019_SonifigrapherSonifiedLight, herrmann_2020_VisualizingSonifyingHow, matsubara_2016_CollaborativeStudyInteractive}, for ordinal data~\cite{adhitya_2011_SonifiedUrbanMasterplan, lindetorp_2021_SonificationEveryoneEverywhere, peng_2023_SirenCreativeExtensible}, and for nominal data~\cite{groppe_2021_SoundDatabasesSonification, hildebrandt_2016_CombiningSonificationVisualization, north_2016_UnderstandingGitHistory}, and now will continue with a discussion of three methods to map the data to the senses, the levels of redundancy.

\subsection{Level of Redundancy}
\label{sec: Level of Redundancy}

\begin{figure}
    \centering
    \includegraphics[width=1\linewidth]{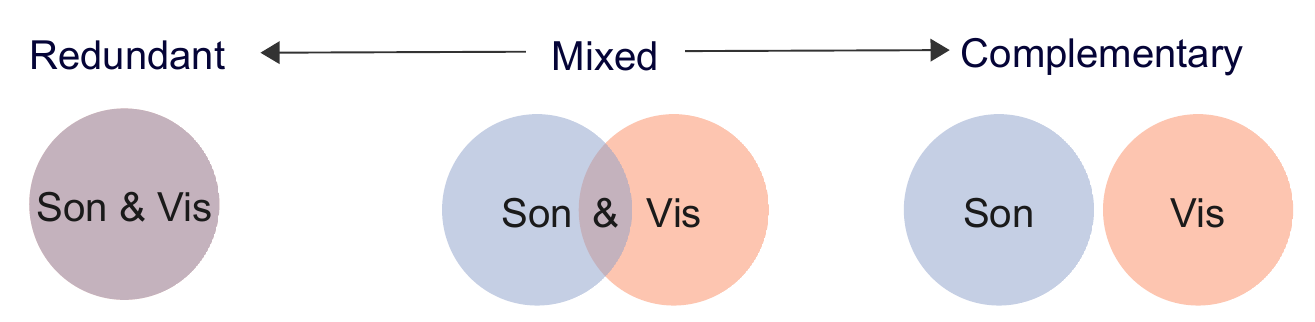}
    \caption{The three levels of mapping-redundancy: Redundant mappings display the same information via visualization and sonification. Complementary mappings display different information to the two senses, and mixed mappings map some information redundantly and others in a complementary way.}
    \label{fig: level of redundancy}
\end{figure}

When combining sonification and visualization, designers have three options -- displayed in \autoref{fig: level of redundancy} -- to distribute the information they want to transport to the senses. They can have the same information represented in both the sonification and the visualization (redundant mappings), they can display one part of the data exclusively visually and another exclusively auditorily (complementary mappings), or they can map some parts of the data redundantly and some complementary (mixed mappings). While this continuum seems to be an intuitive description of audiovisual display idioms, we should distinguish between ``technical redundancy'' and ``communicative redundancy.''

\textbf{Technical redundancy} describes the technical mapping from the data to a visual and auditory representation. If all the displayed information is mapped to channels of both modalities, then the audiovisual display idiom employs a technically redundant mapping. An example is the auditory line graph: We see and hear identical information via spatial position (visualization) and pitch (sonification). In some cases, it is not enough to use a purely technical description of redundancy without incorporating our way of perceiving the different displays as humans. Therefore, we introduce the term "communicative redundancy."

\textbf{Communicative redundancy} describes the fact that technically redundant designs might encode the same information with different perceptual qualities. Hence, we could identify different patterns in data by listening to them than by looking at them. Communicative redundancy will usually be strongly related to technical redundancy, but there are exceptions, such as the combination of a WAV form visualization (line chart) and an audification, such as in the \href{https://www.youtube.com/watch?v=3PJxUPvz9Oo}{Sonification of the Tohoku Earthquake} in Japan in 2011, also described in~\cite{winters_2015_SonificationTohokuEarthquake}. Technically, those two are fully redundant (the same data attributes are displayed both visually and auditorily), but communicatively, they complement each other. 

In the following, when we speak of redundancy, we use the definition of technical redundancy. Communicative redundancy is most likely also dependent on the individual receiver, which is why we would not be able to consistently assign a tag to each case. Nevertheless, when implementing an audiovisual display idiom, a designer should consider its communicative redundancy just as well.

Rönnberg and Johannson~\cite{ronnberg_2016_InteractiveSonificationVisual} present a technically redundant mapping of the density of a \href{https://vimeo.com/247770770}{scatterplot or a parallel coordinates plot} to auditory channels.  Users would be able to hover over the plot and listen to an auditory representation of the density in the data.
It is technically redundant as we are generally able to see the density of the plot by looking at it. In a communicative sense, the design is not redundant, which is why their evaluation shows that the additional sonification helps users identify especially dense areas. The eyes' ability to assess the visual density is supported by the sound of their design.

Dedicated environments that offer the design of audiovisual display idioms frequently employ redundant mappings~\cite{phillips_2019_SonificationWorkstation, riber_2018_PlanethesizerApproachingExoplanet,cantrell_2021_HighchartsSonificationStudio,delavega_2022_SonoUnoWebInnovative,kondak_2017_WebSonificationSandbox,garciariber_2019_SonifigrapherSonifiedLight}. One of them is the \textit{Sonifigrapher} (see \autoref{fig: frameworks}(c)), presented by García~\cite{garciariber_2019_SonifigrapherSonifiedLight}, that combines a line chart and a parameter mapping sonification of the light curves from NASA’s exoplanet archive. The vertical position of a line is mapped to the pitch of a musical sound, essentially ``playing back'' the line from left to right (compare \autoref{fig: platforms}). 

The work presented by Rau et al.~\cite{rau_2015_EnhancingVisualizationMolecular} complementarily enhances the visualization of molecular simulations by using auditory icons and parameter mapping sonification. A user can position a ``virtual microphone'' inside of a 3D-rendered visual representation of a molecule and listen to visually occluded processes.

Temor et al.~\cite{temor_2021_PerceptuallymotivatedSonificationSpatiotemporallydynamic} presented a mixed mapping approach for the audiovisual analysis of computational fluid dynamics tailored towards the prediction of cerebral aneurysm ruptures, hence in the medical context. \href{https://www.youtube.com/playlist?list=PLRbQXqE-XKzdDOqxnXI4yfo3r_cIcrjNl}{Their sonification design} is psychoacoustically motivated in a way that amplifies the differences between different simulations of their spatiotemporally dynamic data. The authors explicitly mention their choice to apply a mixed mapping in their design. They motivate their decision by the observation that the ''\textit{spatiotemporal fluctuations are highlighted in a way that seems to be superior to the presentation of different information to different sensory modalities, which is in line with how we make sense of spatiotemporally-dynamic stimuli in everyday scenarios.}'' We consider this observation highly plausible and inspirational for potential future research and will reflect upon it in the concluding discussion of this STAR.

Maças \& Martins, and Machado present an audiovisual display idiom that displays \href{https://vimeo.com/270078256}{consumption patterns} collected from Portuguese supermarkets over the course of two years. The authors, just in line with Temor et al.~\cite{temor_2021_PerceptuallymotivatedSonificationSpatiotemporallydynamic}, also argue for the employment of a mixed design, their teams having made prior experience with complementary designs that seem to have been less effective.

\noindent
\textbf{Discussion:}
Regarding the level of mapping redundancy, a high-level observation is the following: When designing an audiovisual display idiom, mixed mapping seems to show a special potential. The mapping overlap between the sonification and the visualization seems to help the user perceive an idiom as integrated rather than as two displays existing next to each other. There are different options to design a mixed mapping, out of which a promising one seems to be to synchronize a visual animation with a sonification~\cite{temor_2021_PerceptuallymotivatedSonificationSpatiotemporallydynamic, macdonald_2018_DataDrivenSonificationCFD}. \revized{In a similar manner, synchronizing} the spatial position of the visual display with the direction that a user perceives the sound from can be helpful to perceptually integrate the two stimuli~\cite{chabot_2017_ImmersiveVirtualEnvironment}. \new{An audiovisual analysis idiom that utilizes complementary or mixed mappings enables a designer to choose the employment of sound instead of a second view. Data that would conventionally be made visible via a second view can be represented using sound instead, which can modify the design of the visual view itself. Such situations can become especially apparent whenever the visualization needs to fit on small screens such as on smartwatches.}

\subsection{Evaluation Approaches}
\label{sec: Evaluation Approaches}

The evaluation of designs is a pressing issue in both the sonification and visualization fields. \revized{Audiovisual idioms likewise need to be evaluated}. Rönnberg and Forsell even argue for the standardization of questionnaires that assess the usability of audiovisual representations that could be used in combination with other measurements~\cite{ronnberg_2022_question}.
To study the current practice of evaluation of audiovisual designs, we applied four of the classes of evaluation techniques suggested by Isenberg et al.~\cite{isenberg_2013_systematic}. The four classes are user performance (UP), user experience (UE), algorithmic performance (AP), and qualitative result inspection (QRI).

When evaluations measure how specific features in a visualization or a sonification affect user performance with a system, these approaches belong to the evaluation class UP~\cite{isenberg_2013_systematic}. Controlled experiments using various time measurements and accuracy are typical example methods in this class. Rönnberg and Johansson~\cite{ronnberg_2016_InteractiveSonificationVisual} explored the combination of visualization and sonification, where the users had to identify the visual area with the highest density in visual representations with and without the support of sonification. The authors measured accuracy and response time, and the results showed that the combination of visualization and sonification increased accuracy in comparison with visualization only but also that response times were longer when sonification was used.

The class of UE includes evaluations that focus on subjective feedback on and experiences with a visualization or a sonification~\cite{isenberg_2013_systematic}. Interviews and/or various questionnaires are common evaluation methods. In our corpus of papers, this was a common approach for evaluating the audiovisual representations and was often used in combination with other evaluation approaches. The paper by Ballweg et al.,~\cite{ballweg_2016_InteractiveSonificationStructural} presents a study where users answered a questionnaire about their experience of an audiovisual system for drug design on a 5-point Likert scale. By comparing responses given before and after the study, it could be determined how the system could be integrated into the users’ workflow, if sonification could have a positive effect on solving the task, and in what way the system could be improved. 

Few papers in the systematic literature review used AP as an evaluation approach. Evaluation approaches in this class should contain a quantitative study of the performance or quality of visualization and/or sonification algorithms~\cite{isenberg_2013_systematic}. However, in our corpus of papers, this approach was only found in a few cases and was employed to determine that an audiovisual algorithm could produce sufficient quality rather than determining a certain level of quality or comparing different algorithms. As an example, in the paper by Kariyado et al.~\cite{kariyado_2021_AuralizationThreeDimensionalCellular}, it is stated that the evaluation showed that the system presented in the paper allowed for a default amount of 255 audio sources with any audio drop-outs. 

Evaluations in the class QRI aim to draw conclusions based on qualitative discussions and assessments of audiovisual representations~\cite{isenberg_2013_systematic}. In contrast to UE, these types of evaluations do not involve end users but instead, ask the user to assess the representation for themselves. In the work of Bru et al.~\cite{bru_2023_LineHarpImportanceDriven}, an approach to a combined audiovisual representation based on parallel coordinates or dense line charts is presented. Several different attributes and characteristics in the representations are discussed based on the researchers’ own reflections, but no formal user evaluation has been performed.

\noindent
\textbf{Discussion:}
\revized{We have now demonstrated there are various approaches for evaluating audiovisual representations. The aim of the work and the research questions that are asked determine the class of evaluation chosen.} In some cases, in the corpus of papers, two of these evaluation approaches have been combined to provide a better and more detailed analysis of the outcome of the study findings~\cite{ballweg_2016_InteractiveSonificationStructural,gionfrida_2016_TripleToneSonification,pate_2022_CombiningAudioVisual}. Yet about a third of the studies included in the systematic review did not include an evaluation at all. The absence of evaluation might lead to a situation where promising audiovisual design ideas might be rejected because clear and convincing evidence of their usability, benefits, and function is not presented in a paper, or less useful ideas are overrated and promoted.

\subsection{Target Platforms and Interactivity}
\label{sec: Target Plattforms and Interactivity}

\begin{figure*}
    \centering
    \includegraphics[width=\linewidth]{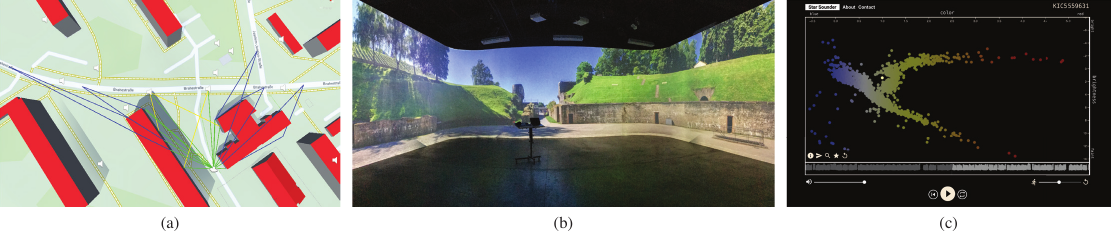}
    \caption{Examples of different target platforms and environments. (a) The combination of VR visualization and sonification for an immersive exploration of noise in urban environments by Berger and Bill~\cite{berger_2019_CombiningVRVisualization}. (b) An immersive virtual environment for audiovisual spatialized data sonification presented by Chabot and Braasch~\cite{chabot_2017_ImmersiveVirtualEnvironment}. (c) A screenshot showing the interactive interface for the sonified Hertzsprung-Russel diagram by Huppenkothen et al.~\cite{huppenkothen_2023_SonifiedHertzsprungRussellDiagram}. All screenshots CC BY 4.0.}
    \label{fig: platforms}
\end{figure*}

Audiovisual display idioms can be displayed on different target platforms and in different environments (see examples in \autoref{fig: platforms}). A typical display for the combination of sonification and visualization is the computer desktop environment, as this is also the most commonly used environment for visualization-only designs. About two-thirds of the designs included in this survey are developed for a desktop environment.

The second largest category of target platforms was physical and/or multi-user environments. This includes dome theaters and planetariums, which are commonly used for audiovisual display idioms toward topics of astronomy (as previously mentioned in \autoref{sec: thematic corpus overview})~\cite{harrison_2022_AudioUniverseTour, russo_2022_5000ExoplanetsListen, elmquist_2021_OpenspaceSonificationComplementing}. Another type of physical environment is a dedicated immersive environment for collaborative data analysis, where several users can take part in the data exploration. One example of such an environment is the \textit{CRAIVE-Lab} (Collaborative-Research Augmented Immersive Virtual Environment Laboratory), where a design visualized and sonified market data of 128 corporations by using the large panoramic display and the 128-channel loudspeaker array of the environment\cite{chabot_2017_ImmersiveVirtualEnvironment} (see \autoref{fig: platforms}(b)).

A low-cost alternative for physical environments is a virtual one, which is most commonly enabled through head-mounted displays. An approach for an audiovisual display idiom in this type of environment is to let the user navigate the dataset from a first-person perspective and use spatialized sonification, which dynamically changes depending on where the user is positioned in the dataset. An example is the design of Berger \& Bill (see \autoref{fig: platforms}(a)), which facilitates an immersive exploration of urban noise standards~\cite{berger_2019_CombiningVRVisualization} by creating a virtual environment of a city where the sonification allows the user to listen to the collected noise levels by navigating the environment.

The use of touch displays and other tangible interfaces for an audiovisual display idiom can allow for a unified integration of sonification and visualization. A recurring approach is to use the visualization as a graphical user interface that enables the sonification upon interaction. Through a touch display, the design by Ferguson et al.~\cite{ferguson_2012_NavigationInteractiveSonifications} enables the user to filter a line graph with multi-touch gestures to select what data should be sonified. As a more analog approach, the system \href{https://www.youtube.com/watch?v=829r3y01XLk}{ \textit{Sonophenology}} by Ness et al.~\cite{ness_2012_Sonophenology} lets the user select areas on a printed color-coded paper map through fiducial markers, which in turn selects what information should be conveyed by the sonification. 

About two-thirds of the papers in the database include some form of interaction. One of the most recurring objectives of the interaction was to make a selection of the dataset that would be sonified~\cite{ferguson_2012_NavigationInteractiveSonifications}. This can include selecting a geographic region to which the dataset is attributed to~\cite{ness_2012_Sonophenology}.
For example, the audiovisual display idiom of Matsubara et al.~\cite{matsubara_2016_CollaborativeStudyInteractive} creates an interactive sonification system for exploration of seismic data through a horizontal and vertical range slider to specify a geographic region. Other forms of interaction include simpler tasks such as browsing sonifications of individual or groups of data objects~\cite{garciariber_2019_SonifigrapherSonifiedLight, lemmon_2023_MappingEmergencyDesigning}. Huppenkothen et al.~\cite{huppenkothen_2023_SonifiedHertzsprungRussellDiagram} created an interactive multimedia version of the Hertzsprung-Russell Diagram, where a user can select individual data entries of the diagram to listen to their sonification. On the other hand, there also exist more complex forms of interactions such as navigating a 3D environment~\cite{berger_2019_CombiningVRVisualization,joliat_2013_SpatializedAnonymousAudio, macdonald_2018_DataDrivenSonificationCFD}. 

Papachristodoulou et al.~\cite{papachristodoulou_2015_AugmentingNavigationComplex} created a design for navigating complex datasets of brain networks, where the sonification conveyed information about different brain regions of the dataset. 
Within the sonification community, it was the introduction of model-based sonification~\cite{hermann2002sonification}, that made user interaction an integral part of the data analysis process. An example of a model-based sonification is the \href{https://pub.uni-bielefeld.de/record/2920473}{\textit{Mode Explorer}} by Yang and Hermann~\cite{yang_2018_InteractiveModeExplorer}, where scratching-interactions with a pencil on a scatter plot enable the user to investigate different modes and their properties through the sonification.

\noindent
\textbf{Discussion:}
When displaying an audiovisual display idiom on a target platform other than a desktop environment, it often creates opportunities for interacting with the system in a novel manner~\cite{elmquist_2021_OpenspaceSonificationComplementing, han_2022_FutureRedVisualizing, ness_2012_Sonophenology, papachristodoulou_2015_AugmentingNavigationComplex}. In the context of data analysis, interaction is also a particularly relevant part of the user experience~\cite{ballweg_2016_InteractiveSonificationStructural, bearman_2011_UsingSoundRepresent, elmquist_2021_OpenspaceSonificationComplementing, han_2022_FutureRedVisualizing}.

\subsection{Users and Goals}
\label{sec: Users and Goals}

Just as with visualization-only designs, audiovisual display idioms are often developed toward specific users and end goals. The most commonly occurring goal for the papers in the database was ``data analysis.'' This could either be to present a dataset to the user 
or letting the user explore one or several datasets with the idiom, as described in \autoref{sec: Purpose}. The benefit of using visualization and sonification in this regard is that the user can gain different perspectives of the data through the two sensory modalities. 
For example, Alonso-Arevalo et al.~\cite{alonso-arevalo_2012_CurveShapeCurvature} created a \href{https://www.youtube.com/@SATINproject/videos}{multimodal interface} for curve shapes and curvature, where the user can evaluate the quality of a three-dimensional shape by using both the visual and auditory, and in this case even their haptic perception. The curve shape was mapped to the fundamental frequency of different carrier sounds, and the idiom offered different sound designs to convey the information. 

The most recurring pair of users and goals in the database was data analysis for domain experts, which most commonly would involve an audiovisual display idiom to convey data in a specific application domain in the natural sciences. Temor et al.~\cite{temor_2021_PerceptuallymotivatedSonificationSpatiotemporallydynamic} and MacDonald et al.~\cite{macdonald_2018_DataDrivenSonificationCFD} created an auditory complement to a scientific visualization of computational fluid dynamics, where \href{https://www.youtube.com/watch?v=UmDvnPjnpV4}{the sonification} aided in understanding the temporal changes in the visual animation.
Gionfrida et al.~\cite{gionfrida_2016_TripleToneSonification} suggest combining the visualization of brain scans (in the context of Alzheimer's dementia research) with a parameter mapping sonification that they call \textit{Triple-Tone Sonification}. The design makes use of the fact that the human ear will perceive the sound of two or more very similar frequencies as "frequency beating," which is also employed in~\cite{malikova_2019_VisualauditoryRepresentationAnalysis}. 
For the purpose of process execution data analysis, Hildebrandt et al.~\cite{hildebrandt_2016_CombiningSonificationVisualization} demonstrated how incorporating sound could enhance the process of identifying anomalies or conducting root cause analysis for irregularities and errors. 
A design for a more general user case was created by North et al.~\cite{north_2016_UnderstandingGitHistory}, which created a \href{https://cse.unl.edu/~myra/artifacts/GitVS/vm/}{an idiom} to convey Git version control data. The temporal nature of the data lent itself well to be sonified by sequentially going through the data, where an earcon is played when a commit occurs, and drum sounds are played whenever a conflict occurs in the data. 

The second most commonly occurring pair of users and goals was public engagement for the general public, where the use of an audiovisual display idiom has the potential to enable higher engagement by conveying information through both the visual and auditory modality. This would commonly be targeted towards popular science topics such as climate change~\cite{han_2022_FutureRedVisualizing, ballora_2015_TwoExamplesSonification} and astronomy~\cite{harrison_2022_AudioUniverseTour, traver_2023_HarmonicesSolarisSonification}.
As previously mentioned in \autoref{sec: Purpose}, the work by Russo et al.~\cite{russo_2022_5000ExoplanetsListen} is an example of a sonification targeted towards the general public for public engagement, where the response of the resulting videos on social media was in part used as a metric for the public engagement of the design.

\noindent
\textbf{Discussion:}
Regarding the successful design of audiovisual idioms that support their users in analyzing data, it is most relevant to include them in the design process. Both the visualization community and the sonification community have individually studied the relevance of including domain experts in their design process~\cite{sedlmair_2012_DesignStudy,goudarzi2017systmematic}, and the same will be necessary for the integration of sonification and visualization. The co-author network shown in \autoref{fig: co-author-net} displays quite many domain experts being collaborators in the surveyed literature, which can be regarded as a promising sign in general.

\section{Survey Data Analyses} \label{Sec: SurveyDataAnalyes}

So far, we have focused on the discussion of the audiovisual idiom designs themselves. In this section, we want to take an even broader perspective on the collected data. We will study existing relationships between different tags along a correlation matrix in \autoref{sec: Correlation Matrix Analysis}, as well as the network of co-authors and its implication for the field in \autoref{sec: co-author network and field}.

\subsection{Correlation Matrix Analysis}
\label{sec: Correlation Matrix Analysis}

\begin{figure*}
    \centering
    \includegraphics[width=1\linewidth]{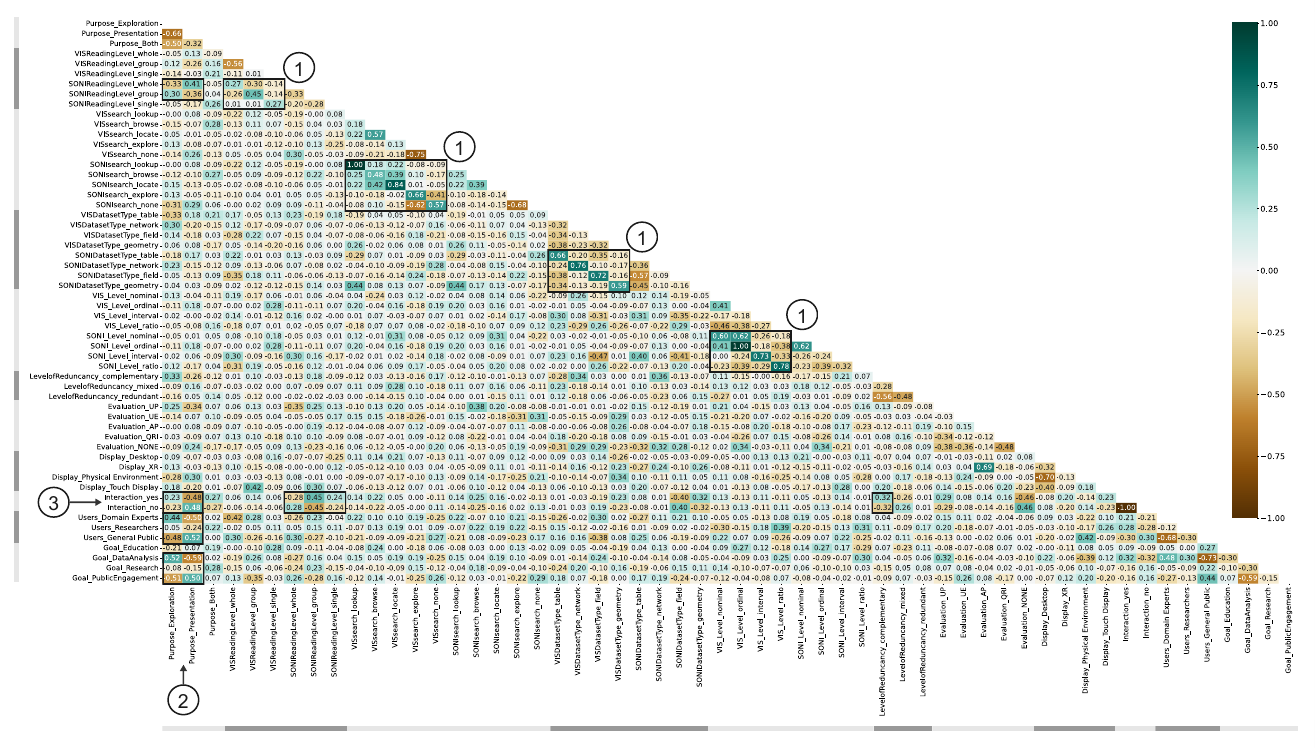}
    \caption{The correlation matrix of a selection of classification tags. A diverging color palette is used with white in the center representing no correlation, teal (up) increasingly positive correlations, and brown (down) increasingly negative correlations. (1) reveals similar tags for visualization and sonification parts; (2) shows correlations between purpose, reading level, interactivity, and intended audience; (3) uncovers relationships with the interaction tag. A high-resolution version of this figure is part of the supplemental material.}
    \label{fig: corrmat}
\end{figure*}

\revized{To understand potential relationships between different tags within our classification, we computed a correlation matrix, shown in \autoref{fig: corrmat}. We want to highlight some of the found correlations in the data out of which most are not surprising, nevertheless, they are not obvious and our data enables us to take such a meta-perspective.}
 
Close to the main diagonal of the correlation matrix, we highlighted four smaller matrices \textbf{(1)}, showing an interesting phenomenon: Whenever the tagging of the papers was done individually for the visualization and the sonification parts, the two were mostly tagged with the same label. Hence, the visualizations and the sonification, in many cases, use the same \textit{reading levels}~\cite{bertin_1983_semiology}, the same \textit{search levels}~\cite{munzner_visualization_2015}, the same \textit{dataset types}, and the same \textit{levels of measurement}. This finding can motivate future research regarding the reading levels and the search levels. Studies could investigate the potential of other distributions of the reading level onto the senses.

Marked with \textbf{(2)}, we highlighted four phenomena concerning the purpose of an audiovisual idiom. Idioms designed to present rather than explore data are likely to also use the reading level ``whole'' for their sonification ($r = 0.41$). On the other hand, idioms that are designed for exploration likely use the group reading level ($r = 0.3$). We see a correlation between the purpose of an idiom and its interactivity, with idioms for exploration more likely to be interactive ($r = 0.23$) and idioms for presentation more likely not to be interactive ($r = - 0.48$). 
Also, domain experts use more idioms for exploration ($r = 0.44$), while the general public is more likely to be `just' presented with data ($r = 0.50$), going hand in hand with their general goals of data analysis ($r = 0.52$) or public engagement ($r = 0.50$).

Two more relationships with the interaction tag emerge from the correlation matrix \textbf{(3)}. A sonification that uses the reading levels of ``group'' or ``single'' is likely to offer user interaction ($r_{group} = 0.45$ and $r_{single} = 0.24$), while a sonification with the reading level of ``whole'' does not usually offer user interaction ($r_{whole} = -0.28$). Furthermore, designs that map data to the senses in a complementary way are more likely to offer interaction to their users ($r = 0.32$).

\revized{Overall, most of the computed correlation values in the correlation matrix are close to zero. Also, some of the existing higher values are likely not significant due to the low amount of data for the respective tags (such as with the AP evaluation, which has been used in only two papers). Therefore, we only discussed three perspectives that have considerable correlation values, are plausible, and that we consider most relevant to the bigger picture.}

\subsection{Co-Author Network and Development Over Time}
\label{sec: co-author network and field}

\begin{figure}
    \centering
    \includegraphics[width=1\linewidth]{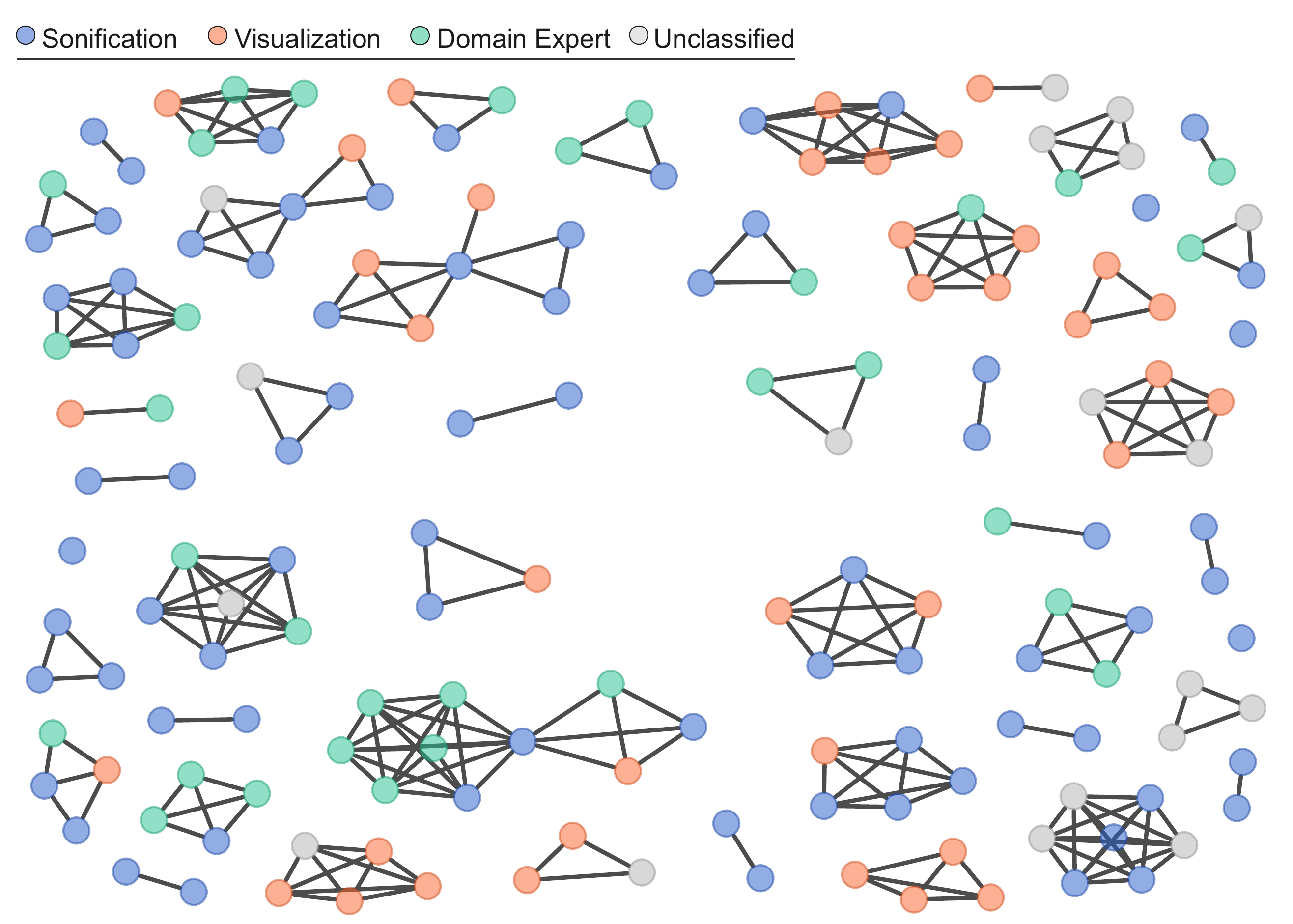}
    \caption{The co-author network, displaying 165 authors of the 57 papers in the database. The links represent papers written by a team of co-authors, and the colors display their primary disciplines. An interactive version of the network can be explored using the Orange workflow, which is part of the supplemental material.}
    \label{fig: co-author-net}
\end{figure}
As part of our meta-analyses, we studied the co-author network within our database\new{, providing a good picture of the state of the research community interested in audiovisual analytics.}
Each node in the co-author network represents an author with a total of 165 nodes (in 48 teams), and a link connects two nodes if the corresponding authors have collaboratively contributed to a publication.
We assigned a color to each node using the primary discipline of the author and classified them, to the best of our abilities, into one of four disciplines: sonification, visualization, domain, and unclassified.
For each author, we considered their main publication focus, particularly around the time of their contribution to the works in our database, along with their background, education, their role and input to the relevant papers.
Authors classified under ``domain'' are recognized as domain experts, and the category ``unclassified'' encompasses scenarios where the author's primary discipline does not fall into any of the other three categories, the author is in the early stages of their career without a defined research focus, or that the information about the author is not available.

The network offers several insights into the structure of the field. The majority of co-author teams are individual teams without interconnections. Only in four cases have authors collaborated with different (groups of) co-authors, usually because one person worked with different collaborators. This phenomenon differs from networks in other fields, such as the one described in a recent STAR on the use of embeddings in visual analytics~\cite{huang_2023_VAEmbeddingsSTARa}, showing a network with clusters up to the size of 75 authors (note that the STAR on embeddings has 122 entries). We interpret the dominance of many disjoint groups as a sign for the rather young field and hope, with this STAR, to contribute to the future development of the field growing closer together. We identified 14 teams or individual authors contributing to \revized{the corpus that consist of sonification experts exclusively. Two of the teams consist only of visualization researchers, and 21 of the teams are collaborations between the communities, including domain experts in many cases.} The inclusion of domain experts and the quite large number of diverse groups also including domain experts can be considered a promising sign for the future development of the field.

Our data also enables us to take a basic temporal perspective on the development of the field. During the years 2011 and 2023, overall, 57 papers were published.
The majority of papers were published in the second half of this time span, pointing towards a generally increasing trend. A drop in publications in the year 2020 is likely a phenomenon related to the COVID-19 pandemic, being dominant worldwide, especially during the year 2020.

\section{Adjacent Topics}%
\label{sec: adjacent}

As mentioned earlier, three distinct topics are adjacent to the scope of our STAR: audiovisual idioms in the context of accessibility, real-time monitoring, and arts. 
Although the three areas are related, they might be better served by a classification system different from the one we used. Still, they are relevant and can be inspirational for our field. The following three sections present work from those adjacent fields that have been curated by our team of co-authors. The sections do not claim completeness or the usage of a systematic approach but are intended as a general introduction for the interested reader.

\subsection{Accessibility}
Even though visualization is one of the most common ways of communicating data, many visualizations are inaccessible to readers with visual impairments. Some studies have suggested the use of sonification to support readers with visual impairments \new{and a recent STAR on accessibility research in visualization identified 16 out of 56 papers that utilize sonification \cite{kim_2021_accessible}}. During the filtering and classification of \revized{our} assembled literature, a category emerged exploring sonification for accessibility of visualization. The literature that was sorted into this category focused on the design and evaluation of sonification to improve the accessibility of visualizations for visually impaired people. As the literature in this category did not explore integrated audiovisual representations for data analysis in general but rather used sonification as a support technique, these are not included in the systematic part of our STAR.

Sonification for accessibility in this way often suggests sonification designs leaning towards auditory graphs. Auditory graphs are the auditory equivalent to visual representations, such as plots, graphs, and charts~\cite{hermann_2011_handbook}, but instead of mapping data properties to line positions or the size of an area, they are mapped to various auditory parameters such as pitch or loudness. Auditory graphs are suggested to be useful as sensory substitution and as a means of presenting information when line of sight is not possible~\cite{stockman2005auditory}, and using sound to represent data makes data analysis more accessible~\cite{walker2010universal}. To give the reader a brief introduction to this category, here are a few recent studies that have explored sonification for accessibility in visualization. 

One study explored the accessibility of data visualizations presenting data to blind and visually impaired users~\cite{fan2023accessibility}. It was found that sonification might make discerning trends in the data more accessible, but also that the lack of experience in using sonification could lead to misinterpretations of the presented data. The study was conducted using screen readers, and the users often used alt-text information to support and validate what was heard through the sonification. Another study explored the use of natural sounds mapped to data visualizations, bar charts, and line charts~\cite{hoque2023accessible}. The reason for using natural sounds was to support users without musical training in understanding the sonification and auditory graph, and it was found that these natural sounds could support the understanding of categorical data and were most useful for users without musical training. Infographics, the combination of visualization and text information, has been explored with an auditory-only approach~\cite{holloway2022infosonics}. An interactive approach, infosonic, was explored to facilitate accessibility of data to blind and low-vision users, using spoken introduction and annotation, and non-speech sonification. The study shows that the sonification approach supported understanding and forming a mental image of the data. 

Often, accessibility studies tend to blend sonification, i.e.,~non-speech sounds, with generated speech or screen readers. This makes the interpretation of the results from these studies somewhat challenging. The findings,~i.e., the experiences and the understanding of the sonification, might be due to the sonification itself or depending on other auditory stimuli used. Also, the visual representation, the user interaction, and the user interface are of importance to the study outcome. In visual data analysis, interaction is essential for exploring the data~\cite{ceneda2019review}, and similarly, for sonification to be a useful tool, dynamic human interaction is necessary~\cite{Hunt:2004:IIS}. 
Sonification for accessibility of visualization for visually impaired users is an exciting approach and some of these findings could suggest interesting approaches also for sighted users in supporting visual perception with sonification~\cite{ronnberg2019musical,ronnberg2019sonification}, or reducing cognitive load~\cite{zagermann2016measuring} on the visual modality by sonification (see for example~\cite{mousavi1995reducing}).

\subsection{Monitoring}
A comprehensive overview or a systematic literature review of audiovisual process monitoring is far beyond the scope of this report. This section is intended to provide some insights into this research field and to distinguish the various scientific communities involved.

The combination of visual and auditory displays for process monitoring purposes has been well-established in real-world applications and interfaces. The audiovisual representation of Morse code for an exchange of information serves as an early example of a multimodal monitoring display and interface. In supervising and controlling the states of various devices and processes, audiovisual interfaces combine the advantages of both modes: volatile alerting sound signals enhance situation awareness, while visual cues provide additional information about the situation, enabling users to take action. Implementations of audiovisual interfaces are ubiquitous in both professional and day-life environments. For instance, consider the warning beep, which can be heard when the temperature outside your car drops below 4°C, alerting you to the possibility of slippery road conditions. A quick glance at the visual display will provide you with the exact temperature and the opportunity to decide whether to take action or not. 

Besides the approaches to audiovisual data analysis described in the previous sections of this report, audiovisual monitoring interfaces play an integral role in various domains, including air traffic control~\cite{elmquist_sonair_2023}, control rooms~\cite{sirkka_design_2022,hoferlin_auditory_2012}, anesthesia~\cite{andrade_augmenting_2021,roche_anesthesia_2022}, neurology~\cite{lin_2018_VisualizationSonificationLongTerm}, dermoscopy~\cite{walker_dermoscopy_2019}, surgery~\cite{ziemer_three-dimensional_2023}, network monitoring~\cite{worrall_polymedia_2019,axon_sonification_2021}, automotive~\cite{jakus_user_2015,xin_effectiveness_2022}, to name a few.

During the literature search for this STAR, several publications were labeled with the tag ``monitoring.'' After closer investigation, which involved excluding multiple publications written by identical authors on very similar topics, as well as publications addressing topics of accessibility and artistic media installations, only a relatively small representative subset of 14 findings qualifies for audiovisual process monitoring within the search criteria of this STAR. 

The reasons for this rather limited number of retrieved articles in this category are manifold. \revized{Firstly, most research on monitoring applications used for auditory feedback of human behavior in the fields of sports, therapy, or rehabilitation focuses on the design, impact, and evaluation of interfaces that utilize the non-intrusive attributes of auditory displays. Hence, they enable users to perceive information without any interfering actions, such as adjusting their head or body position to view a visual display~\cite{schaffert_review_2019,van_rheden_sonification_2020,maculewicz_investigation_2016}.} While some of these interfaces include visualizations at a low scale level, they hardly qualify as audiovisual monitoring devices in terms of providing a balanced contribution from both modalities.

In industrial and surveillance contexts, factors such as cognitive workload, perception organization, situational awareness, alarm fatigue, and deafness play crucial roles in the successful implementation of infrastructures. However, these attributes are discussed in other scientific communities, including Human Factors and Applied Ergonomics, as well as several areas of medicine and health, particularly in anesthesia, rehabilitation, sports, and cognitive psychology. In addition to performance comparisons between auditory, visual, and audiovisual modalities~\cite{audry_congruent_2019}, particular interest is given to the dual-task paradigm, which evaluates, for instance, the ability to identify an event requiring action in a secondary domain while simultaneously performing a primary task~\cite{lu_supporting_2013,hildebrandt_continuous_2016,tardieu_sonification_2015,nees_prototype_2014}.

(Non-)systematic literature reviews have been published in several of the mentioned areas~\cite{wang_interactive_2017,kanev_sonification_2019,schlosser_head-worn_2021,hildebrandt_server_2015}. To the best knowledge of the authors, no comprehensive STAR covering the entire spectrum of audiovisual or related multimodal monitoring applications and approaches has been conducted thus far. However, there are some fundamental publications treating the specific attributes and criteria for research and development in the field, such as~\cite{iber_2020_handbook,johannsen_auditory_2004,watson_designing_2007}.

\subsection{Arts}
\label{sec: art}
During the classification of the corpus, an ``art'' category emerged, defining projects combining visualization and sonification for artistic purposes. We understand the emergence of art practices that intertwine sonification and visualization not only as inspiring for designers and audiences alike but also relevant to the broader discourse around data representation towards a better, more efficient, and engaging human-data relationship.

Working with the literature, we understood artistic endeavors call for a different kind of classification than idioms intended for audiovisual data analysis. We decided not to include such contributions in the systematic part of this STAR and excluded papers where authors made their intentions of an artistic contribution explicit in the paper.

Nevertheless, to give the reader a brief introduction to the field, we present representative cases of an emerging art category that could form the basis of future research on the integration of sonification and visualization. In addition to the artistic cases that emerged from our literature search, we use four more sources to curate a collection of representative cases for artistic endeavors: The \href{https://sonification.design/}{Data Sonification Archive} (DSA), the \href{https://direct.mit.edu/comj}{Computer Music Journal}, the \href{https://direct.mit.edu/lmj}{Leonardo Music Journal}, and the \href{https://ars.electronica.art/festival/en/}{Ars Electronica Festival}. The Data Sonification Archive is an online crowd-sourced collection of data sonification projects. Out of 455 cases currently hosted in the DSA, two of the co-authors of this report recently categorized 139 cases as ``art'', i.e., as having their primary purpose in creating and delivering an artistic experience to an audience~\cite{lindborg_2023_ClimateDataSonification}. A preliminary search on some of the major venues for academic artistic publications -- Computer Music Journal and Leonardo Music Journal -- accounts for 182 and 118 cases, respectively, in line with the volume of art-related cases in the Data Sonification Archive. A search of the archive of the renowned Ars Electronica Festival on projects that translate data into an audiovisual idiom returns 32 cases, again, in line with the 23 projects currently hosted on the DSA.

Typically, data-driven artworks that use an audiovisual design address a non-expert audience and can take different forms, such as live performances, multimedia installations, video productions, or web experiences where the sonification and visualization of the same dataset are combined. The DSA classification highlights recurring topics such as climate change and/or the representation of environmental data, as well as individual and collective interaction in the urban space, internet, and social media. The predominance of socially relevant topics shows that the emerging art category presents borderline characteristics with the adjacent category (in the DSA classification system, see~\cite{lindborg_2023_ClimateDataSonification}) of ``public engagement''. This latter category identifies projects that, although being often presented in the form of an artistic experience, have explicitly and primarily the goal of increasing awareness -- even fostering activism -- among the general public on a specific phenomenon.

\providecommand{\idiom}[1]{\emph{#1}}

This is the case for projects that present data related to climate change, such as \idiom{Klima|Anlage---Performing climate data}~\cite{gross-vogt_2019_klima}, one of the cases identified as artistic in our literature search. In this work, climate data and model predictions between 1950 and 2100 can be chosen by the listener for twelve selected regions of the world and interactively sonified and visualized in an immersive installation environment with the goal of contributing to ``the urgent need to inform the general public about climate change''.

In \idiom{Too Blue}~\cite{foo_2015_tooblue}, hosted on the DSA, the author created an audiovisual experience (in the form of a video) where music generated by tracking the land loss in coastal Louisiana over 78 years due to man-made levees, drilling and dredging for oil and gas, and climate change is combined with an aerial color-coded visual map that shows the increase in the ``blues'' colored areas, corresponding to the sea, in the coastal land of Louisiana. In \idiom{Heat and the Heartbeat of the City}~\cite{polli_2006_heat}, the urgency to communicate climate change to the broader audience combines with a reflection on the urban space through the simultaneous sonification and visualization of actual and projected data on increased temperature in New York City. In \idiom{aqua\_forensic}~\cite{sebjanic_2018_aqua} data on invisible pollutants in the Danube river's water, and the effect on its ecosystem, are collected, sonified, and visualized `to present the results of this scientific research to the wide audience in a poetic and artistic way.' Other prominent social issues such as personal data collection and privacy in the age of social media are at the center of multi-awarded art projects such as \idiom{Digital Violence}~\cite{forensic_2021_violence} where sonification is explicitly used to increase the audience's engagement with the visualized dataset on the issue of illegal government surveillance practices. From the collective to the individual experience, projects such as \idiom{Orbuculum}~\cite{ylmaz_2023_orbuculum}, and \idiom{Deep Sync}~\cite{kiesenhofer_2023_deepsync} invite the public to interactively engage with personal psychological and physiological dimensions to create an immersive auditory and visual experience. In a more-than-human perspective, works such as \idiom{Spider Web Sonification}~\cite{su_2020_sonification} and \idiom{Biota Beats}~\cite{kim_2020_SonifyingDataHuman} shift the focus from humans to other species in an effort to support the public engagement with science.

Data art can stimulate an ``artistic affectivization'' that can contribute to building shared perspectives within a community~\cite{buening_2022_BetweenFactandFabrication}. The engagement of the audience with the phenomenon behind the data, however, also implies an information and knowledge transfer~\cite{masud_2010_FromDatatoKnowledge}, i.e., the public not only has to connect emotionally with the data but also has to understand their meaning. The emergence of a category of artistic projects that combine sonification and visualization with the explicit intent to activate the audience on a socially relevant issue shows that multimodal approaches have the potential to generate impact, not only for education and awareness but also for stimulating action.

\section{Concluding Discussion and Future Work}
\label{sec: discussion}

In this STAR, we have provided an overview of the field of audiovisual idiom design tailored toward data exploration and presentation.
We have used a variety of perspectives to classify the existing literature, and, in \autoref{sec: adjacent}, we offered insights into three topics that are adjacent to our report and can be highly inspirational for future developments.
For our categorizations, we also provided brief discussions or reflections as part of the individual subsections \ref{sec: Purpose} to \ref{sec: Users and Goals}. 

\new{Before we adopt a global viewpoint and discuss broader insights, challenges, and exciting future research opportunities, we now want to mention some of the key insights and results from previous sections.
Overall, our corpus holds a quite diverse field of combined designs, employing visual idioms such as scatter plots, maps, and networks, as well as sonification techniques such as parameter mapping sonification, audification, and earcons.}

\new{The analysis of employed reading levels \cite{bertin_1983_semiology} of the different designs revealed an imbalance between the two modalities. The visualization part of the identified audiovisual idioms mostly works at higher reading levels than their sonification counterparts. This phenomenon suggests that many designs provide an overview of the data using their visualization, while the sonification supports the analysis of details.}

\new{Such design decisions are also related to the level of redundancy an audiovisual idiom employs, and mixed approaches, which represent parts of the data redundantly and other parts in a complementary manner, seem to show special potential. Such a selective alignment of mapping between the sonification and visualization might assist us in perceiving an idiom as integrated rather than as two separate displays coexisting next to each other.}

\new{Furthermore, the option to use sonification as part of an idiom's design can directly influence the visual view as well. Thinking of idioms with ``multiple views,'' one of the views could, for example, be covered by an ``auditory view.'' For visualization designers, such considerations are especially relevant when space is limited, as is the case, for example, on smartwatches~\cite{lee_2021_MobileDataVisualization}.}

\noindent
\textbf{The culture of providing supplemental material:}
A thriving future for the field of audiovisual idiom design also depends on the communities' culture when providing demos for their designs as supplemental material. Out of 57 articles in our surveyed literature, only 29 provided a demo that is still available today. Five provided a demo that is no longer available, and 23 provided no demo at all. Online repositories such as \href{https://zenodo.org/}{Zenodo}, hosted by CERN\new{, the platform \href{https://osf.io/}{OSF}, maintained by the Center for Open Science}, or long-term archives such as \href{https://phaidra.univie.ac.at/}{Phaidra}, hosted by the University of Vienna, allow the upload of videos and even the assignment of DOIs. While making sound available in a paper used to be an explicit challenge to the sonification field, it is not the case anymore. An additional visual representation of the sound can also be included in the paper, using a spectrogram of the sound itself or an iconic representation of it.

\noindent
\textbf{Evaluating audiovisual idioms:} 
It is often challenging to conduct systematic, well-informed, and, above all, properly performed and reported evaluations of audiovisual idioms, especially when the goal is to compare different studies.
As mentioned earlier, Rönnberg and Forsell suggested the standardization of questionnaires tailored to the field in order to increase the comparability between different studies~\cite{ronnberg_2022_question}.
Establishing this comparability is particularly difficult because the two senses are used simultaneously in an audiovisual display, and the evaluation methods must adapt to this characteristic of the design.
Specifically, the existing methods focus on several aspects.
One is user performance, which studies quantitative metrics such as task completion times and precision of responses.
Another is user experience, which focuses more on qualitative metrics such as user engagement and memorability.
From the data we collected in our STAR, we can draw high-level conclusions regarding the type of evaluations that have been used within the corpus of literature.

\revized{Among the papers that included a user evaluation (21 of the 57 papers), most reported a subjectively stated benefit of sonification in perceiving and interpreting the visualization.
Some subjective ratings and feedback also suggested that the use of sonification was engaging (six papers considered this~\cite{han_2022_FutureRedVisualizing,russo_2022_5000ExoplanetsListen,elmquist_2021_OpenspaceSonificationComplementing,kariyado_2021_AuralizationThreeDimensionalCellular,ronnberg_2021_SonificationConveyingData,du_2018_ExploringRoleSound}), could provide emotional content (mainly two papers~\cite{russo_2022_5000ExoplanetsListen,ronnberg_2021_SonificationConveyingData}), was experienced as interesting (five papers~\cite{han_2022_FutureRedVisualizing,russo_2022_5000ExoplanetsListen,elmquist_2021_OpenspaceSonificationComplementing,ronnberg_2021_SonificationConveyingData,du_2018_ExploringRoleSound}), as well as pleasant (four papers~\cite{russo_2022_5000ExoplanetsListen,elmquist_2021_OpenspaceSonificationComplementing,ronnberg_2021_SonificationConveyingData,du_2018_ExploringRoleSound}), and in perceived improved quality of visual information (six papers~\cite{pate_2022_CombiningAudioVisual,elmquist_2021_OpenspaceSonificationComplementing,kariyado_2021_AuralizationThreeDimensionalCellular,fitzpatrick_2018_StreamSegregationUtilizing,ronnberg_2016_InteractiveSonificationVisual,bearman_2012_UsingSoundRepresent}). 
Other papers also reported subjective and objective measurements of improved interpretation and higher understanding of data when sonification and visualization were used in combination (16 papers~\cite{pate_2022_CombiningAudioVisual,cantrell_2021_HighchartsSonificationStudio,elmquist_2021_OpenspaceSonificationComplementing,du_2018_ExploringRoleSound,fitzpatrick_2018_StreamSegregationUtilizing,gune_2018_GraphicallyHearingEnhancing,yang_2018_InteractiveModeExplorer,ballweg_2016_InteractiveSonificationStructural,gionfrida_2016_TripleToneSonification,north_2016_UnderstandingGitHistory,ronnberg_2016_InteractiveSonificationVisual,papachristodoulou_2015_AugmentingNavigationComplex,papachristodoulou_2014_SonificationLargeDatasets,alonso-arevalo_2012_CurveShapeCurvature,bearman_2012_UsingSoundRepresent,bearman_2011_UsingSoundRepresent}). This was shown in terms of higher scores or increased accuracy when sonification was used. However, this increase in accuracy was reported to be related to increased task completion times (see, for example, in~\cite{ronnberg_2016_InteractiveSonificationVisual}). Similar effects have also been shown in other studies not included in this systematic review~\cite{metatla2016sonification,ronnberg2019musical,ronnberg2019sonification}. The case might be that the use of the additional information provided by sonification that makes the increased accuracy possible takes a longer time to process and assess. While traditionally increased task completion times are considered a disadvantage, in some contexts, this can be seen as positive. Slowing down the user, resulting in them spending more time with their data, studying it from more and from different perspectives (including the auditory perspective), might reveal structures or patterns in a dataset that would have stayed silent and unseen if analyzed with a conventional unimodal display~\cite{hullman_2011_benefitting}.}

\revized{These results indicate that there is, in general, a benefit of combining visualization and sonification for data exploration and analysis, regardless of the specific visualization and/or sonification design.}
\new{Nevertheless, we consider it a major task for the future research of our community to systematically study what kinds of benefits the integration of sonification and visualization can provide.}

\noindent
\textbf{The potential of unconventional task distribution:}
The data from our STAR shows that reading levels~\cite{bertin_1983_semiology} are not equally distributed to the two senses: The vast majority -- 50 out of 57 papers used the reading level ``whole'' for their visualization (see \autoref{tab:tag-definitions}), while only about half of the papers use the same reading level for their sonification parts. This observation might be directly related to the way audiovisual idioms distribute the tasks of overview and detail to the two senses. The currently more popular distribution seems to be to use visualization to display an overview of the data, while sonification is employed to provide details. While this distribution seems like an intuitive choice, future research should study the potential of switching the roles between sonification and visualization. It is our daily lives and the way we experience the world around us that suggests sonifying an overview and visualizing details. Our auditory sense constantly screens our 360° environment while our visual sense actively focuses on details in our environment. The field of auditory process monitoring makes use of exactly this situation when alerting us of a status, regardless of our current visual focus. Nevertheless, we identified only three cases where the reading level for the sonification was ``whole'' and the one for the visualization was ``group''~\cite{delavega_2022_SonoUnoWebInnovative, ferguson_2012_NavigationInteractiveSonifications, yang_2018_InteractiveModeExplorer}.

A second phenomenon we observe in the data is that many audiovisual display idioms employ the same search level~\cite{munzner_visualization_2015} (see \autoref{fig: corrmat}), and most of them employ the search level of ``explore'' (see \autoref{tab:tag-definitions}). Again, it seems like a fruitful future endeavor to study the opportunities of designs that explicitly break such a pattern. 

\noindent
\textbf{The potential of integrating sonification and scientific visualization:}
Scientific visualization typically displays inherently spatial phenomena and is often used to represent data that varies over time. The inherently spatio-temporal quality of such displays constitutes their potential to be integrated with sonification, as sonification is a display technique tailored toward temporal data structures~\cite{guttman_2005_hearing,enge_2023_unified}. 
We relate this argument back to the real-world example from our introduction, where the combination of our visual and auditory senses offers a better understanding of a phenomenon than each sense alone: Rain falling outside of a closed window. Opening the window to not only see but also hear the sound of the rain gives us a better estimation of the amount and intensity of the rainfall. Abstractly, in such a situation, we perceive a spatio-temporal phenomenon. It is most plausible that adding sound to a time series of scientific visualization can increase its holistic interpretability in an ecologically valid manner~\cite{neuhoff_ecological_2004}. This phenomenon is shown in two publications in the database, in the work by MacDonald et al.~\cite{macdonald_2018_DataDrivenSonificationCFD}, and the work by Temor et al.~\cite{temor_2021_PerceptuallymotivatedSonificationSpatiotemporallydynamic}. Similarly, these ideas also apply to animations in information visualization, where temporal aspects of the data might be visualized and sonified.

\noindent
\textbf{Design frameworks:}
When it comes to the possibilities of designing audiovisual display idioms, most researchers, as well as most domain experts, will not be able to develop their own designs. Interdisciplinary knowledge bridging visualization, sonification, interactive design, and human perception is necessary to design effective, engaging, and re-usable audiovisual display idioms. Therefore, we cannot expect a domain expert to be able to quickly draft a prototype the same way they may be able to do using established visualization-only frameworks such as \textit{matplotlib}~\cite{Hunter_2007_matplotlib}.

In our database, we identified several design frameworks tackling this  challenge~\cite{phillips_2019_SonificationWorkstation,peng_2023_SirenCreativeExtensible,lindetorp_2021_SonificationEveryoneEverywhere,delavega_2022_SonoUnoWebInnovative,kondak_2017_WebSonificationSandbox,cantrell_2021_HighchartsSonificationStudio}. Their primary focus is the design of the sonification part of the idiom, which is why we cannot speak of truly balanced contributions to both senses. A promising endeavor to enable truly balanced designs, with both the visualization and the sonification being equally well-informed, might be to not design them using one but two separate frameworks, each specialized for its purpose. In this regard, the recent work by Reinsch and Hermann~\cite{hermann_2021_Sc3nbPythonSuperColliderInterfacea, reinsch_2022_InteractingSonificationsMesonic, ReinschHermann-ICAD2023-Sonecules} is promising, as it provides a sonification design framework embedded in the \textit{Python} environment and is conceptually inspired by visualization design tools such as \textit{matplotlib}. With both the sonification design and the visualization design happening in the same environment, in this case, for example, a \textit{Jupyter Notebook}, both can be developed with the required quality to meet their individual standards. 
Regarding the introduction of users to a new design environment, be it for visualization, sonification, or their combination, effective ``onboarding'' is a pressing issue. In the visualization community, the topic has been studied in recent years~\cite{stoiber_2023_designb, stoiber_2023_visualization}, and the same will be necessary for sonification and audiovisual display idioms. Again, when it comes to sonification as a technique to display data, the prior experience of domain experts or the general public is low. This lack of experience can, in the short term, only be met with carefully designed onboarding processes.

\noindent
\textbf{Adjacent topics as inspiration to the field:}
Our brief look into the adjacent fields of monitoring, accessibility, and arts shows a special potential for inspiration for our communities' future work. The field of monitoring offers established evaluation methods that could potentially be adopted to support the evaluation of audiovisual idioms tailored toward exploratory data analysis. During our unstructured investigation into the field, we also identified the potential of a systematic STAR as a future contribution to the field of monitoring. 

Also, the adjacent field of accessibility can inspire future research, where understanding the analysis patterns of visually impaired individuals can inform design decisions for idioms tailored towards sighted people~\cite{walker2010universal}. While the field of accessibility can be informative to the field of audiovisual idioms, the same holds true the other way around. Especially the designs from our STAR that use redundant mapping to both the visual and the auditory senses can inspire future designs for visually impaired users. Furthermore, such designs have the potential to foster successful collaborative data analysis involving both sighted and blind users. 

The identified artistic contributions show great potential to be inspirational to the field of audiovisual idiom design as well. In the future, effort should be made by the research community to develop a specific framework for the evaluation and analysis of audiovisual idioms at the crossing of art and public engagement. Such efforts need to explore how specific design strategies (e.g., interactivity and embodiment, as used in many of the cases presented in \autoref{sec: art}) can be combined to support both affect and sense-making. 
Again, a systematic STAR dedicated to the artistic perspective would be a highly timely contribution to the community.

\noindent
\textbf{An audiovisual analytics community:}
With this state-of-the-art report, we hope to contribute to the future establishment of systematic research on audiovisual display idiom design. The current development concerning the studied co-author network and the temporal development of publication numbers in the field is promising. We hope to reach researchers from both the visualization and sonification communities, inspiring them for future interdisciplinary collaborations, and to \textit{open their ears \revized{and} take a look}.


\section*{Supplemental Material}

As supplemental material to this STAR, we provide at \url{https://phaidra.fhstp.ac.at/o:5541}:

\begin{itemize}
    \item \textsf{corpus.bib} and \textsf{corpus.rdf}: publication metadata of the surveyed literature in BibTeX and Zotero RDF format.
    \item \textsf{corpus-tagging.csv}: a table holding the surveyed literature and all of the used tags. Interested readers may use the table to identify articles that fall into specific combinations of classes.
    \item \textsf{authors.csv}: a table holding the names of the authors in our STAR corpus and their primary discipline. Interested readers may use it to identify potential future collaborators.
    \item \textsf{corrmat.pdf}:  a high-resolution version of \autoref{fig: corrmat}.
    \item \textsf{authornet.ows}: the Orange Data Mining workflow file, used to generate the co-author network in \autoref{fig: co-author-net} (incl.\ the used color scheme file \textsf{authornet\_colorscheme.colors}).
\end{itemize}

In addition, the corpus of relevant papers with tags, publication metadata, and full text of open access work is available as a \textit{Zotero} library at \url{https://www.zotero.org/groups/integrationsonificationvisualization/items}.

\section*{Acknowledgments}
We would like to thank PerMagnus Lindborg, School of Creative Media, City University of Hong Kong, for valuable feedback. This research was funded in part by the Austrian Science Fund (FWF): 10.55776/P33531 and the Gesellschaft f\"ur Forschungsf\"orderung Nieder\"osterreich (GFF) SC20-006, as well as by the Knut and Alice Wallenberg Foundation (grant KAW 2019.0024). %

\bibliographystyle{eg-alpha-doi}  
\bibliography{ref-sonivis-star}                   

\end{document}